\documentclass[showpacs,prl,superscriptaddress,nofootinbib, twocolumn]{revtex4-1}
\usepackage[english]{babel}
\usepackage[utf8]{inputenc}
\usepackage[babel]{csquotes}
\usepackage[T1]{fontenc}
\usepackage{amsthm}
\usepackage{amsmath}
\usepackage{amsfonts}
\usepackage{mathrsfs}
\usepackage{amssymb}
\usepackage{mathtools}
\usepackage{graphicx}
\usepackage{subfigure}
\usepackage{geometry}
\usepackage{braket}
\usepackage{sidecap}
\usepackage{float}
\usepackage{bm}
\usepackage{lipsum}

\theoremstyle{remark}

\begin{document}

\title{Resilience for stochastic systems interacting via a quasi-degenerate network.}

\author{Sara Nicoletti}
\affiliation{Universit\`{a} degli Studi di Firenze, Dipartimento di Fisica e Astronomia,
CSDC and INFN, via G. Sansone 1, 50019 Sesto Fiorentino, Italy}
\affiliation{Dipartimento di Ingegneria dell'Informazione, Universit\`{a} di Firenze,
Via S. Marta 3, 50139 Florence, Italy}
\author{Duccio Fanelli}
\affiliation{Universit\`{a} degli Studi di Firenze, Dipartimento di Fisica e Astronomia,
CSDC and INFN, via G. Sansone 1, 50019 Sesto Fiorentino, Italy}
\author{Niccol\`{o} Zagli}
\affiliation{Imperial College London, South Kensington Campus, London SW7 2AZ}
\author{Malbor Asllani}
\affiliation{MACSI, Department of Mathematics and Statistics, University of Limerick, 
Limerick V94 T9PX Ireland}
\author{Giorgio Battistelli}
\affiliation{Dipartimento di Ingegneria dell'Informazione, Universit\`{a} di Firenze,
Via S. Marta 3, 50139 Florence, Italy}
\author{Timoteo Carletti}
\affiliation{naXys, Namur Institute for Complex Systems, University of Namur, Belgium}
\author{Luigi Chisci}
\affiliation{Dipartimento di Ingegneria dell'Informazione, Universit\`{a} di Firenze,
Via S. Marta 3, 50139 Florence, Italy}
\author{Giacomo Innocenti}
\affiliation{Dipartimento di Ingegneria dell'Informazione, Universit\`{a} di Firenze,
Via S. Marta 3, 50139 Florence, Italy}
\author{Roberto Livi}
\affiliation{Universit\`{a} degli Studi di Firenze, Dipartimento di Fisica e Astronomia,
CSDC and INFN, via G. Sansone 1, 50019 Sesto Fiorentino, Italy}

\begin{abstract}
...
\end{abstract}


\begin{abstract}
A stochastic reaction-diffusion model is studied on a networked support. In each patch of the network two species are assumed to interact following a non-normal reaction scheme. 
When the interaction unit is replicated on a directed linear lattice, noise gets amplified via a self-consistent process which we trace back to the degenerate spectrum 
of the embedding support. The same phenomenon holds when the system is bound to explore a quasi degenerate network. In this case, the eigenvalues of the Laplacian operator, which governs species diffusion,  accumulate over a limited portion of the complex plane. The larger the network, the more pronounced the amplification. Beyond a critical network size, a system deemed deterministically stable, hence resilient, may turn unstable, yielding seemingly regular patterns in the concentration amount. Non-normality and quasi-degenerate networks may therefore amplify the inherent stochasticity, and so
contribute to altering the perception of resilience, as quantified via conventional deterministic methods.
\end{abstract}
\maketitle
\textbf{Models of interacting populations are of paramount importance in a broad range of applications of interdisciplinary breath. 
Beyond the simplified arena of deterministic approaches, stochastic effects play a role of paramount importance and might yield 
a large of plethora of non trivial behaviors. Furthermore, to account for the inherent complexity of the existing interactions, the inspected model are embedded on a network architecture. In this paper, we show how a non-normal reaction model coupled to a directed, quasi-degenerate, network can drive a resonant amplification of the noisy component of the dynamics. This observation, that we here substantiate analytically, calls for a revised concept of resilience, the ability of a system to oppose external disturbances.}

\section{Introduction}

Resilience represents the inherent ability of a given system to oppose external disturbances and eventually recover the unperturbed state. The concept of resilience is particularly relevant to ecology \cite{resilience1}. Here, perturbations of sufficient magnitude may force the system beyond the stability threshold of a reference equilibrium. When the threshold is breached, recovery is not possible and the system under scrutiny steers towards an alternative attractor, distinct from the original one.  Resilience, namely the capacity of a system to withstand changes in its environment, plays a role of paramount importance for a large plethora of applications, beyond ecology and ranging from climate change to material science, via information security and energy development \cite{resilience2}. 

To grasp the mathematical essence of the phenomenon, one can customarily rely on a straightforward linear stability analysis of the governing dynamical, supposedly deterministic, equations. The eigenvalues of the Jacobian matrix, evaluated at equilibrium, conveys information on the associated stability. If the largest real part of the eigenvalues is negative, the examined deterministic system is deemed stable, against tiny disturbances \cite{murray,strogatz}. To state it differently, the system is able to regain its deputed equilibrium, by exponentially damping the imposed perturbation. However, a linearly stable equilibrium can be made unstable, through non-linearities,  by a sufficiently large perturbation amount: this occurs for instance when the enforced disturbance takes the system outside the basin of attraction, i.e. the set of initial conditions leading to long-time behavior that approaches the attractor \cite{menck}. In the following, we shall focus on sufficiently small perturbations, so that the linear stability holds true. A transient growth can, however, take place, at short times, before the perturbation fades eventually away, as established by the spectrum of the Jacobian matrix. This short time amplification is instigated by the non-normal character of the interaction scheme and may trigger the system unstable, also when the eigenvalues of the Jacobian display a negative real part \cite{trefethen,teo1}. The elemental ability of a non-normal system to prompt an initial rise of the associated norm, can be made perpetual by an enduring stochastic drive \cite{zagli,sara}. Taken altogether, these evidences call for a revised notion of resilience, for systems driven by non-normal coupling and shaked by stochastic forcing. 

We shall be in particular interested in interacting multi-species models, diffusively coupled via a networked arrangement \cite{othmer1,othmer2,Mikhailov,Nakao,ACPSF,HNM,ABCFP,Kouv,PFMC,CBF,PLFC,LFC,teo2}. The investigated systems are assumed to hold an homogeneous equilibrium. For a specific selection of the involved parameters, the homogeneous fixed point can turn unstable upon injection of a non homogeneous perturbation, which activates the diffusion component. The ensuing symmetry breaking instability, as signaled by the so called dispersion relation, constitutes the natural generalization of the celebrated Turing instability to reaction-diffusion systems hosted on a complex network.  In \cite{muolo}, we showed that non-normal, hence asymmetric networks may drive a deterministic system unstable, also if this latter is predicted stable under the linear stability analysis. As we shall here prove, the effect is definitely more remarkable when the non-normal system is made inherently stochastic \cite{gardiner,vankampen}. At variance with the analysis in \cite{biancalani}, it is the non normality of the network to yield the self-consistent amplification of the noisy component, a resonant mechanism which prevents the resilient recovery. To elaborate along these lines, we will consider a generic reaction-diffusion model defined on a directed one dimensional lattice. The degenerate spectrum of the Laplacian that governs the diffusive exchanges between adjacent nodes sits at the root of the generalized class of instability that we shall here address. Making the directed lattice longer, i.e. adding successive nodes to the one-dimensional chain, allows for the perturbation to grow in potency, and for the system to eventually cross the boundaries of stability \cite{clement}. A similar scenario is met when the hosting network is assumed quasi degenerate, meaning that the Laplacian eigenvalues are densely packed within a limited portion of the complex plane. 

Many real networks, from different realms of investigations ranging from biology (neuronal, proteins, genetic) to ecology (foodwebs), via sociology (communication, citations) and transport (airlines), have been reported to possess a pronounced degree of non-normality \cite{teo1}. In particular, it was shown that their adjacency matrix is almost triangular (when properly re-organizing the indexing of the nodes) which, in spectral terms, implies enhancing the probability of yielding a degenerate spectrum for the associated Laplacian matrix. Networks that display a triangular adjacency matrix are known as directed acyclic graphs (DAG).

The paper is organized as follows. In the next section, we shall introduce our reference setting, a reaction-diffusion system anchored on a directed one dimensional lattice.
We will in particular show that a self-consistent amplification of the noisy component of the dynamics is produced, when successively incrementing the number of nodes that form the chain. This bears non trivial consequences in terms of  resilience, displayed by the system under scrutiny. Moving from this preliminary information, in Section II we will modify the lattice structure by accommodating for, uniform and random, return loops. This breaks the degeneracy that arises from the directed lattice topology. The coherent amplification of the stochastic drive is however persistent as long as the spectrum is close to degenerate. Random directed acyclic graphs with quasi degenerate spectrum can be also created, which make reaction-diffusion systems equally prone to the stochastic driven instability. This topic is discussed in Section III.  Finally in Section IV we sum up and draw our conclusion.

\section{Reaction-diffusion dynamics on a directed lattice}

We begin by considering the coupled evolution of two species which are bound to diffuse on a (directed) network.  Introduce the index $i = 1,...,\Omega$ to identify the $\Omega$ nodes of the collection and denote by 
$\phi_i$  and $\psi_i$ the concentration of the species on the $i$-th node. The local (on site) reactive dynamics is respectively governed by the non-linear functions $f(\phi_i,\psi_i)$ and $g(\phi_i,\psi_i)$. 
The structure of the underlying network is specified by the adjacency matrix $\mathbf{A}$: $A_{ij}$ is different from zero if a weighted link exists which connects node $j$ to node $i$. The species can relocate across the 
network traveling the available edges and subject to standard Fickean diffusion. With reference to species  $\phi_i$, the net flux at node $i$ reads $D_{\phi} \sum_{j=1}^{\Omega} A_{ij} (\phi_j-\phi_i)$, where $D_{\phi}$ stands for the diffusion coefficient and the sum is restricted to the subset of nodes $j$ for which $A_{ij} \ne 0$. Furthermore, we shall assume that the dynamics on each node gets perturbed by an additive noise component of amplitude $\sigma_i$. In formulae the system under investigation can be cast as: 
\begin{subequations}
\begin{align}
\frac{d}{dt} \phi_i &= f\left(\phi_i,\psi_i\right)+D_{\phi}\sum_{j=1}^\Omega \Delta_{ij} \phi_j + \sigma_i (\mu_{\phi})_i  \label{eqn: eq di partenza1} \\
\frac{d}{dt} \psi_i &= g\left(\phi_i,\psi_i\right)+D_{\psi}\sum_{j=1}^\Omega \Delta_{ij} \psi_j + \sigma_i (\mu_{\psi})_i\label{eqn: eq di partenza2}
\end{align}
\end{subequations}
where  $\Delta_{ij}=A_{ij}-k_i\delta_{ij}$ stands for the discrete Laplacian and  $k_i = \sum_j A_{ij}$  is the incoming connectivity. Here, $(\mu_{\phi})_i$ and $(\mu_{\psi})_i$ are Gaussian random variables with zero mean and correlations
$\langle (\mu_{\phi})_i(t) (\mu_{\phi})_j (t')  \rangle = \langle (\mu_{\psi})_i(t) (\mu_{\psi})_j (t')  \rangle = \delta_{ij} \delta(t-t')$ and $\langle (\mu_{\phi})_i(t) (\mu_{\psi})_j (t')  \rangle =0$. 

\begin{figure}[ht!]
\includegraphics[scale=0.25]{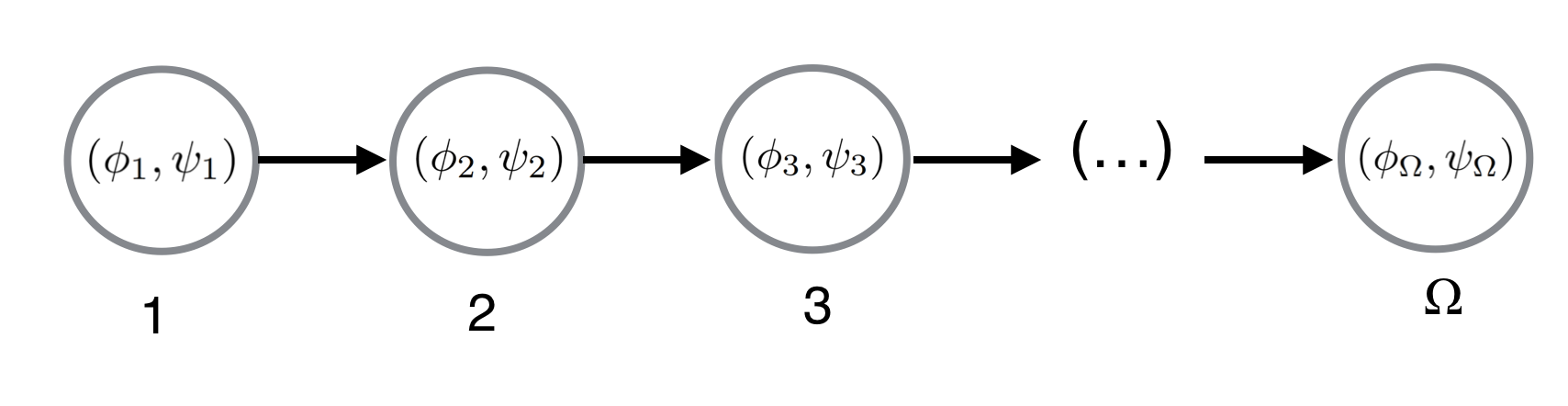}
\caption{The scheme of the model is illustrated. Two populations,  respectively denoted by $\phi$ and $\psi$, are distributed on a collection of $\Omega$ nodes. The nodes are arranged so as to form a one dimensional lattice subject to unidirectional couplings, as outlined in the scheme. The weight of the link between adjacent nodes is set to one. The local, on site, interaction among species is ruled by generic non linear functions of the concentration amount.}
\label{fig1_scheme}
\end{figure}

In this section, we will assume a directed lattice as the underlying network. A schematic layout of the system is depicted in Fig. \ref{fig1_scheme}. The $\Omega \times \Omega$
 Laplacian matrix associated to the directed lattice reads:
\begin{equation}
\mathbf{\Delta}=
\begin{pmatrix}
0 & 0 & \dots \\
1 & -1 & 0 & \dots \\
0 & 1 & -1 & 0 & \dots \\
\vdots & \ddots & \ddots & \ddots  \\
\end{pmatrix}
\end{equation}
and admits a degenerate spectrum, an observation that will become crucial for what it follows. More precisely, the eigenvalues of the Laplacian operators are $\Lambda^{(1)}=0$, with multiplicity $1$, and  $\Lambda^{(2)}=-1$, with multiplicity 
$\Omega-1$. Notice that the eigenvalues are real,  even though the  Laplacian matrix is asymmetric. 

\subsection{The deterministic limit ($\sigma_i=0$).}
To continue with the analysis, we will assume that the deterministic analogue of system \eqref{eqn: eq di partenza1}-\eqref{eqn: eq di partenza2} (obtained when setting $\sigma_i =0$, $\forall i$) admits a homogeneous fixed point and label this latter $(\phi^*,\psi^*)$. In practice,  $\phi_i=\phi^*$ and $\psi_i=\psi^*$, $\forall i$, is an equilibrium solution of the model equations. Furthermore, we will assume that the aforementioned fixed point is stable against homogeneous perturbation, a working assumption which can be formally quantified by considering the associated Jacobian matrix:
\begin{equation}
\mathbf{J}=
\begin{pmatrix}
f_{\phi} & f_{\psi}\\
g_{\phi} & g_{\psi}\
\end{pmatrix}
\end{equation}
where $f_{\phi}$ stands for the partial derivative of $f(\phi, \psi)$ with respect to $\phi$, evaluated at the fixed point $(\phi^*,\psi^*)$. Similar definitions hold for $f_{\psi}, g_{\phi}$ and $g_{\psi}$. The homogeneous fixed point is stable provided $tr(J)=f_{\phi}+g_{\psi}<0$ and $det(J)=f_{\phi} g_{\psi} - f_{\psi} g_{\phi}>0$. Here,   $tr(...)$ and $det(...)$  denote, respectively, the trace and the determinant. 
When the above inequalities are met, the largest real part of the eigenvalues of the Jacobian matrix $J$ is negative, pointing to asymptotic stability. Peculiar behaviors however arise when the spectrum of $J$ is degenerate, namely when the eigenvalues come in identical pair. 

In this case, the solution to the linear problem ruled by matrix $J$ contains a secular term, which, depending on the initial conditions, might yield a counterintuitive growth of the perturbation, at short time. For long enough time the exponential damping takes it over and the system relaxes back to the stable equilibrium (or, equivalently, the imposed perturbation is damped away). Short time transients can also be found in stable linear systems, ruled by non-normal matrices. A matrix is said non-normal, if it does not commute with its adjoint. For the case at hand, $J$ is assumed to be real. Hence, taking the adjoint is identical to considering the transpose of the matrix. In formulae, $J$ is non-normal, provided  $[J, J^T]=J J^T - J^T J \ne 0$, where the apex $T$ identifies the transpose operation. Assume that the non-normal matrix $J$ is stable, hence its eigenvalues have negative real parts. Consider then the Hermitian part of $J$, a symmetric matrix defined as $H(J)=(J+J^T)/2$. If the largest eigenvalue of $H$ is positive, then the linear system governed by $J$ can display a short time growth, for a specific range of initial conditions. Systems characterized by stable Jacobian matrix, with an associated unstable Hermitian part, are termed reactive.
In the following we shall consider a reactive two components system, namely a two species model that possesses the elemental ability to grow the imposed perturbation at short times, also when deemed stable. This latter ability will be considerably augmented by replicating such fundamental unit on the directed chain, and so engendering a secular behavior which eventually  stems from the degenerate structure of the associated Jacobian. 

Heading in this direction we shall first make sure that the examined system is stable when formulated in its spatially extended variant, namely when the two species dynamical system is mirrored on a large set of nodes, coupled diffusively via unidirectional links. A non homogeneous perturbation can be in fact imposed which activates the diffusion component and consequently turns, under specific conditions, the homogenous solution unstable. The subtle interplay between diffusion and reaction weaken therefore
the resilience of the system, by opposing its ability to fight external disturbances and eventually regain the unperturbed (homogeneous) state. 
The conditions for the onset of the diffusion driven instability are obtained via a linear stability analysis, that we shall hereafter revisit for the case at hand. The analysis yields a dispersion relation which bears information on the outbreak of the instability. This latter constitutes a straightforward generalization of the celebrated Turing mechanism to the case of a discrete, possibly directed support.  

To further elaborate along this axis, we focus on the deterministic version of system 
 \eqref{eqn: eq di partenza1}-\eqref{eqn: eq di partenza2} and impose a, supposedly small, non homogeneous perturbation of the homogenous equilibrium. In formulae, we set:
 \begin{eqnarray}
\phi_i&=&\phi^*+\xi_i \\ \nonumber
\psi_i&=&\psi^*+\eta_i
\label{perturb}
\end{eqnarray}
where $\xi_i$ and $\eta_i$ stand for the imposed perturbation. In the following we will label $\boldsymbol{\zeta}=\left(\xi_1,\eta_1,...,\xi_\Omega,\eta_\Omega \right)$ the vector which characterizes the fluctuations around the fixed point. Insert now the above condition in the governing equation and linearize around the fixed point, assuming the perturbation to be small. This readily yields a $2 \Omega \times 2 \Omega$ linear system in the variable $\boldsymbol{\zeta}$:
\begin{equation}
\frac{d}{dt} \boldsymbol{\zeta} = \mathcal{J} \boldsymbol{\zeta}
\label{linear_syst}
\end{equation}
 where 
\begin{equation}
\label{generalizedLapl}
\mathcal{J}=
    \begin{pmatrix}
    \mathbf{J} & \mathbf{0} & \dots \\
    \mathbf{D} & \mathbf{J}-\mathbf{D} & \mathbf{0} & \dots \\
    \mathbf{0} & \mathbf{D} & \mathbf{J}-\mathbf{D} \\
    \vdots & \ddots & \ddots & \ddots
    \end{pmatrix}
\end{equation}
where   $\mathbf{D}=\begin{pmatrix} D_{\phi} & 0\\ 0 & D_{\psi}\ \end{pmatrix}$ is the diagonal diffusion matrix. The spectrum of the generalized Jacobian matrix $\mathcal{J}$ convey information on the asymptotic fate of the imposed perturbation. If the eigenvalues display negative real parts, then the perturbation is bound to fade away, at sufficiently large times. The system recovers hence the unperturbed homogeneous configuration.  Conversely, the perturbation grows when the eigenvalues possess a positive real part. A Turing-like instability sets in and the system evolves towards a different, non homogeneous attractor. 

To compute the eigenvalues $\lambda$ of matrix $\mathcal{J}$, we label $\varepsilon_i=\mathbf{J}+\mathbf{D}\Delta_{ii}$ for $i=1,\dots,\Omega$. Then, the characteristic polynomial of $\mathcal{J}$ reads:
\begin{eqnarray}
0&=&\det(\mathcal{J}-\lambda I)=\prod_{i=1}^{N}\det(\varepsilon_i-\lambda I) \\
&=& \det(\varepsilon_1-\lambda I) \left[ \det(\varepsilon_2-\lambda I) \right]^{\Omega-1}
\end{eqnarray}
where $I$ stands for the $2 \times 2$ identity matrix and
\begin{equation}
    \det(\varepsilon_1-\lambda I)= \lambda^2 -tr(\mathbf{J}) \lambda + det (\mathbf{J}) 
\end{equation}
\begin{eqnarray}
\nonumber
    \det(\varepsilon_2-\lambda I)&=&\lambda^2-tr(\mathbf{J'})\lambda+\det(\mathbf{J'})
\end{eqnarray}
where we have introduced the matrix $\mathbf{J'}=\mathbf{J}-\mathbf{D}$.
The spectrum of the generalized Jacobian is hence degenerate if $\Omega>2$ and the multiplicity of the degeneracy of the spectrum grows with $\Omega$, 
the number of nodes that compose the lattice. In formulae:
\begin{equation}
    \lambda_{1,2}=\frac{tr(\mathbf{J})\pm\sqrt{tr^2(\mathbf{J})-4\det (\mathbf{J})}}{2}
\end{equation}
and 
\begin{equation}
\lambda_{3,4} = \frac{tr(\mathbf{J'})\pm\sqrt{tr^2(\mathbf{J'})-4\det(\mathbf{J'})}}{2}
\end{equation}
with degeneracy $\Omega-1$. The stability of the fixed point $(\phi^*,\psi^*)$ to external non homogeneous perturbation is hence determined by $\left(\lambda_{re}\right)_{max}$, the largest real part of the above eigenvalues. \\

For $\Omega>2$, the solution of the linear system (\ref{linear_syst}) can be cast in a closed analytical form, by invoking the concept of generalized eigenvectors. Denote with $\boldsymbol{v_0}$ and 
 $\boldsymbol{w_0}$ the ordinary eigenvectors associated to the non degenerate eigenvalues $\lambda_1$ and $\lambda_2$. The ordinary eigenvectors $\boldsymbol{v_1}$ and $\boldsymbol{w_1}$, respectively associated to  $\lambda_3$ and $\lambda_4$, are non degenerate (i.e. the geometric multiplicity is one). This can be proven for a generic dimension $2\Omega$ of the matrix $\mathcal{J}$ due to the simple block structure of the matrix. We then introduce 
 the generalized eigenvectors associated to $\lambda_3$, $\lambda_4$ as:
\begin{align}
(\mathcal{J}-\lambda_3 I)^ i\boldsymbol{v_{i+1}} &=\boldsymbol{v_i}, \quad i=1,\dots,\Omega-2 \\
(\mathcal{J}-\lambda_4 I)^i \boldsymbol{w_{i+1}} &=\boldsymbol{w_i}, \quad i=1,\dots,\Omega-2. 
\end{align}
The solution of (\ref{linear_syst}) reads therefore:
\begin{multline}
\boldsymbol{\zeta}=c_0e^{\lambda_1t}\boldsymbol{v_0}+d_0e^{\lambda_2t}\boldsymbol{w_0}\\
+[c_1\boldsymbol{v_1}+c_2(\boldsymbol{v_1}t+\boldsymbol{v_2})+c_3(\boldsymbol{v_1}t^2+\boldsymbol{v_2}t+\boldsymbol{v_3})\\
+\dots+c_{\Omega-1}(\boldsymbol{v_1}t^{\Omega-2}+\boldsymbol{v_2}t^{\Omega-3}+\dots+\boldsymbol{v_{\Omega-1}})]e^{\lambda_3t}\\
+[d_1\boldsymbol{w_1}+d_2(\boldsymbol{w_1}t+\boldsymbol{w_2})+d_3(\boldsymbol{w_1}t^2+\boldsymbol{w_2}t+\boldsymbol{w_3})+\dots\\
+d_{\Omega-1}(\boldsymbol{w_1}t^{\Omega-2}+\boldsymbol{w_2}t^{\Omega-3}+\dots+\boldsymbol{w_{\Omega-1}})]e^{\lambda_4t}.
\end{multline}

Secular terms, which bear the imprint of the spectrum degeneracy, may boost the short time amplification of the norm of the imposed perturbation. Remind that the transient growth occurs for $\left(\lambda_{re}\right)_{max}<0$, i.e. when the perturbation is bound to fade away asymptotically. Interestingly, the degree of the polynomial increases with the lattice size. This implies in turns that the short time amplification of a stable system could progressively grow, as the number of lattice nodes get increased. 

\begin{figure}[ht!]
\includegraphics[scale=0.14]{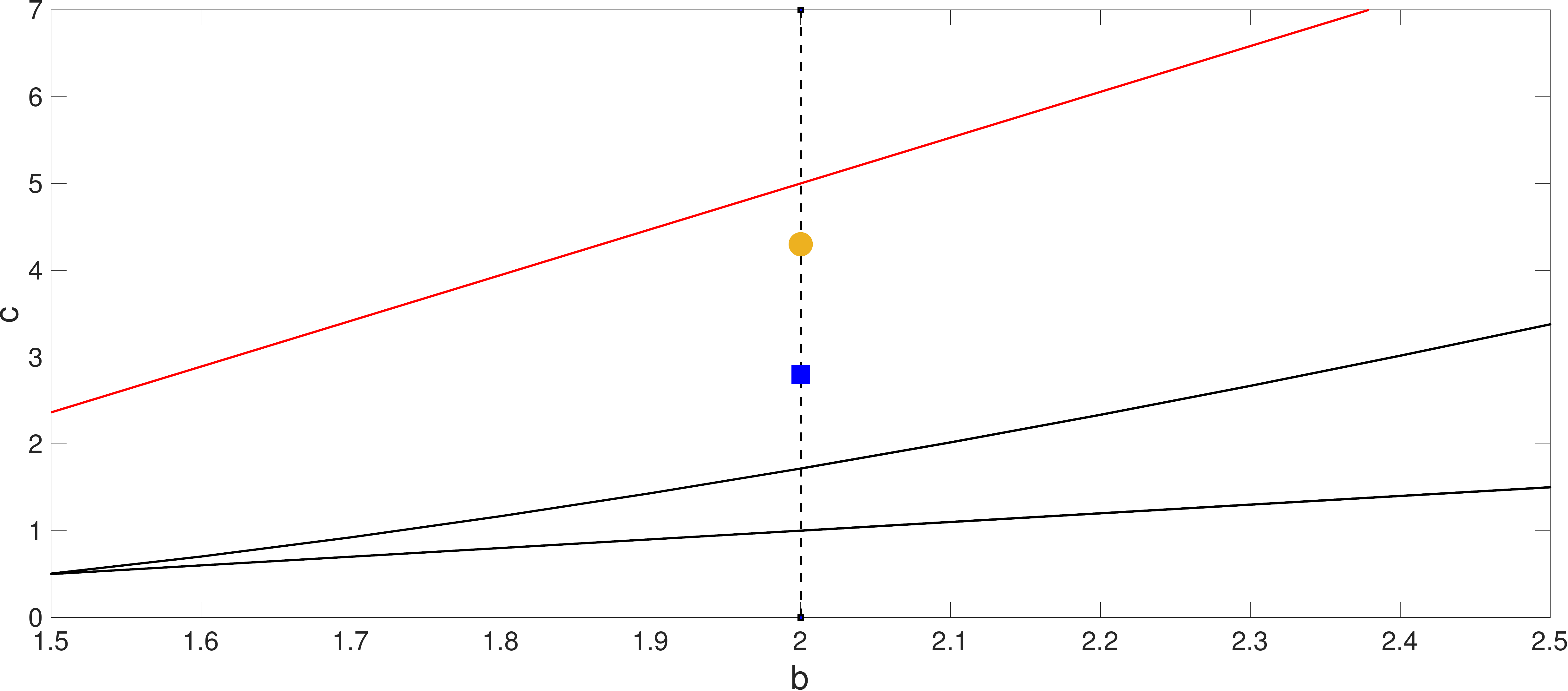}
\caption{The region of instability is depicted in the parameter plane $(b,c)$. This is the portion of the plane contained in between the two black solid lines. The symbols refer to two distinct operating points, positioned outside the domain of deterministic instability. The two points are located at $b=2$, and have respectively $c=2.8$ and $c=4.3$. The red line delimits the 
upper boundary of the parameters region where the stochastic amplification can eventually take place, as discussed in the remaining part of the section. Here, 
$D_{\phi}=1$ and $D_{\psi}=10$.}
\label{fig2}
\end{figure}

\begin{figure}
\begin{tabular}{cc}
\includegraphics[scale=0.25]{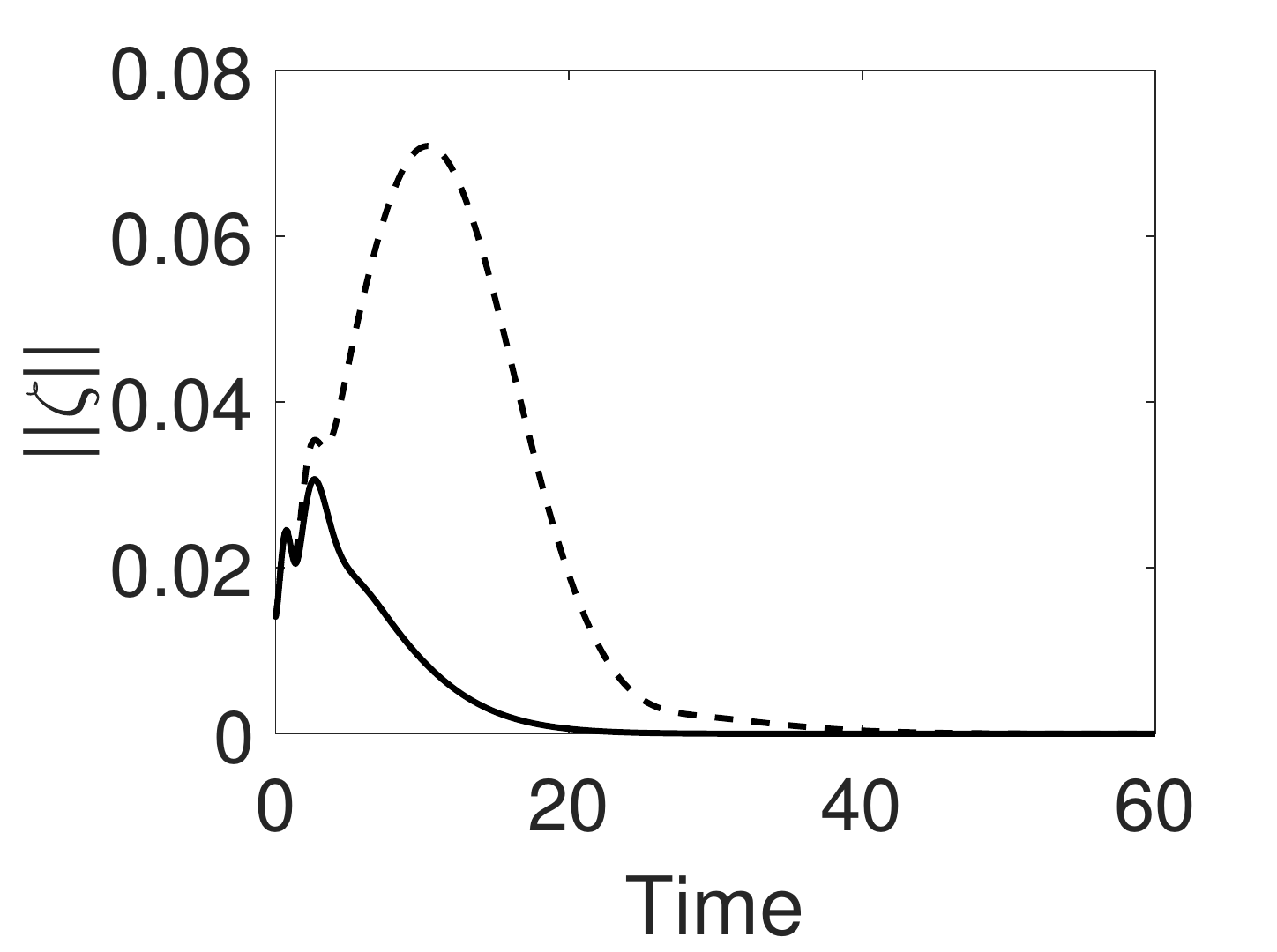} & 
\includegraphics[scale=0.25]{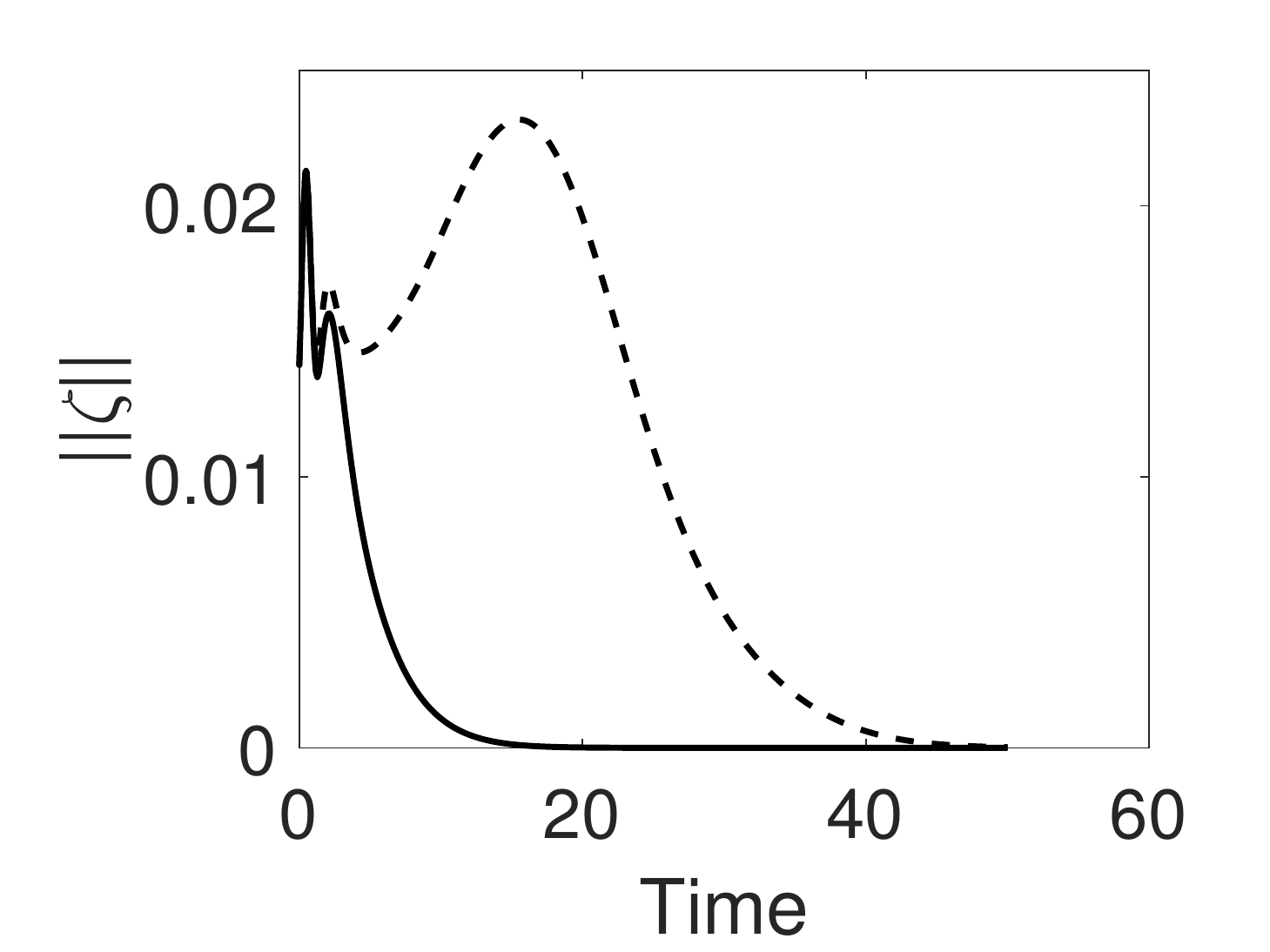} \\
(a) & ( b)
\end{tabular}
\caption{The evolution of the norm of the perturbation is displayed for different choices of the reaction parameter $c$, corresponding to the two points selected in Fig. \ref{fig2} (at $b=2$). (a) Here $c=2.8$, the lower point (blue square) in Fig  \ref{fig2}. The solid line refers to $\Omega=5$ while the dashed line is obtained for $\Omega=10$. 
(b) Here $c=4.3$, the upper point (orange circle) in Fig  \ref{fig2}. The solid line stands for $\Omega=5$ while the dashed line refers  to $\Omega=25$. Notice that the amplification gets more pronounced the closer the working point is to the deterministic transition line. Also, the peak of the norm against time shifts towards the right as the power of the leading secular terms is increased.
Here,  $D_{\phi}= 1$ and $D_{\psi}=10$.}
\label{fig3}
\end{figure}

\begin{figure}
\includegraphics[scale=0.14]{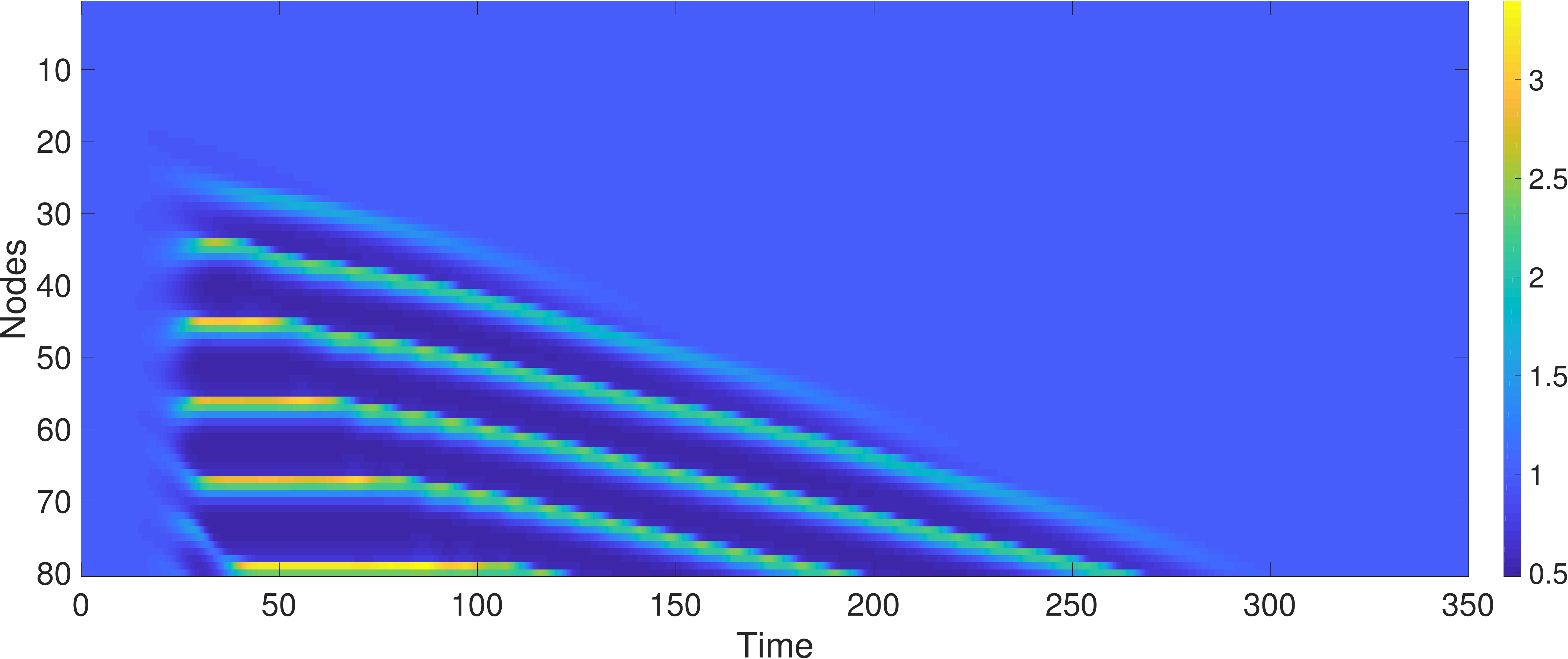} \\
(a) \\
\includegraphics[scale=0.14]{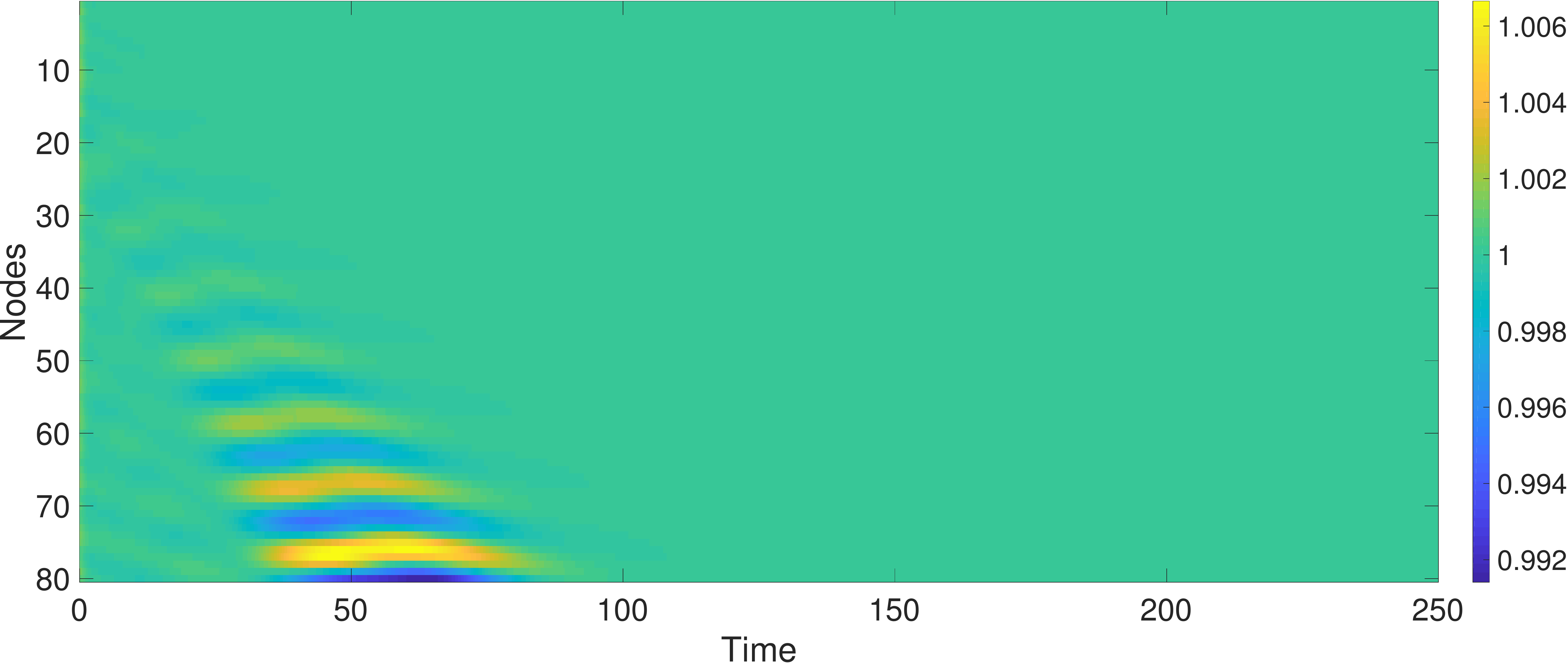} \\
( b)
\caption{The time evolution of species $\phi$ is displayed with an appropriate color code, on different nodes of the chain and against time. The Brusselator reaction scheme is assumed.
Diffusion constants are set to $D_{\phi}= 1$ and $D_{\psi}=10$.  Panel (a) refers to the values of $b$ and $c$ associated to the blue square in Fig. \ref{fig2}, while panel (b) follows the parameters attributed by assuming the orange circle as the operating point.}
\label{fig4}
\end{figure}

Without loss of generality, and to test the implication of the above reasoning, we shall hereafter assume  the Brusselator model, as a reference reaction scheme. 
The Brusselator is a paradigmatic testbed for nonlinear
dynamics, and it is often invoked  in the literature as
a representative model of self-organisation, synchronisation
and pattern formation. Our choice amounts to setting
$f\left(\phi_i,\psi_i\right)=1-(b+1) \phi_i + c \phi_i^2 \psi_i$ and $g\left(\phi_i,\psi_i\right)=b \phi_i-c \phi_i^2 \psi_i$
where $b$ and $c$ stand for positive parameters. The system admits
a trivial homogeneous  fixed point for $\left(\phi_i,\psi_i\right) = (1, b/c)$. This latter is stable to homogeneous perturbation provided  $c > b-1$. We can then isolate in the parameters plane  $(b,c)$ the domain where $\left(\lambda_{re}\right)_{max}>0$, or stated differently, the region that corresponds to a generalized Turing instability. The result of the analysis is displayed in Fig. \ref{fig2}, for a specific choice of the diffusion parameters, with $D_{\phi}>D_{\psi}$: the region of Turing instability falls inside the black solid lines (the red line follows the analysis of the stochastic analogue of the model, as we will explain in the following). The symbols identify two distinct operating points, positioned outside the region of deterministic instability. Working in this setting, after a short time transient,  the perturbations get exponentially damped and the system relaxes back to its homogeneous equilibrium. The effect of the short time amplification should get more visible for increasing values of $\Omega$, the lattice size and the close the operating point is to the threshold of instability. This scenario is confirmed by inspection of Fig \ref{fig3} where the norm of the perturbation is plotted against time, for the chosen parameters value.  To further elaborate on this observation we display in Fig  \ref{fig4} the density of species $\phi$,  on different nodes of the chain, against time. Panel (a) refers to the choice of parameters that corresponds to the blue square in Fig.  \ref{fig2}, while panel (b) follows for the parameters associated to the orange circle. By making the chain longer, one better appreciates the initial growth of the perturbation which materialize in transient patterns. For sufficiently long time, the patterns disappear and the system converges back to the unperturbed homogeneous fixed point. The time for equilibration grows with the size of the lattice, i.e. with the number of nodes $\Omega$. Transient patterns are more pronounced and persistent, the closer to the region of deterministic instability.  


\subsection{The stochastic evolution ($\sigma_i \ne 0$).}

We shall here move on to consider the effect produced by a perpetual noise. This amounts to assuming  $\sigma_i \ne 0$ in \eqref{eqn: eq di partenza1}-\eqref{eqn: eq di partenza2}. For the sake of simplicity we will set in the following $\sigma_i=\sigma$. We anticipate however that our conclusions still hold when accounting for an arbitrary degree of heterogeneity in the 
strength of the noise. By linearizing the governing dynamical system around the fixed point yields a set of linear Langevin equations:  
\begin{equation}\label{eq: Langevin_LIN}
\frac{d}{d\tau}\zeta_i=\left(\mathcal{J}\boldsymbol{\zeta}\right)_i+\hat{\lambda}_i
\end{equation}
where $\hat{\boldsymbol{\lambda}}$ is a $2 \Omega$ vector of random Gaussian entries with  zero mean, $<\hat{\boldsymbol{\lambda}}>=0$, and correlation given by
$<\hat{\lambda}_i(\tau)\hat{\lambda}_j(\tau')>=\sigma^2\delta_{ij}\delta(\tau-\tau')\equiv \sigma^2\delta(\tau-\tau')$.

\begin{figure}[ht!]
\includegraphics[scale=0.2]{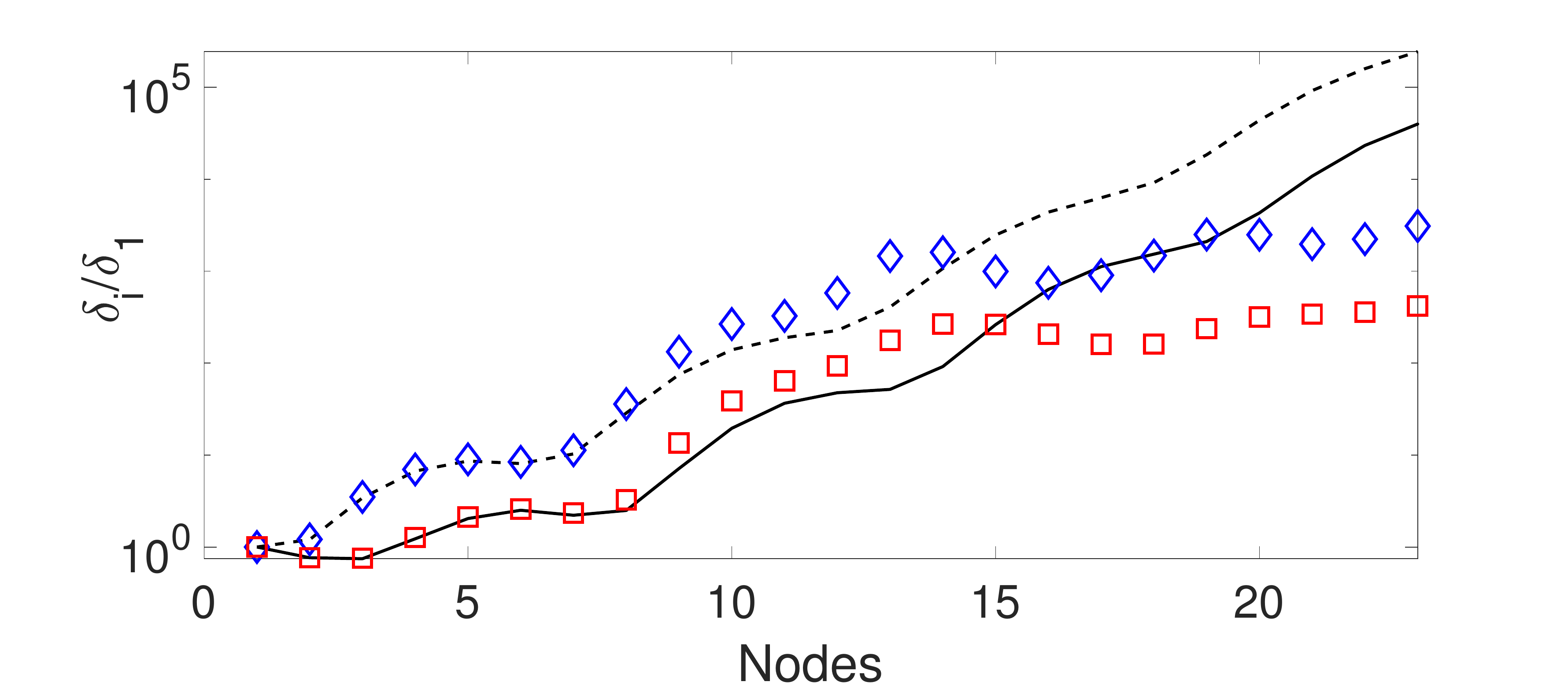}
\caption{The phenomenon of noise driven amplification is illustrated:   $\delta_i/\delta_1$ is plotted against the node's number along the chain. The solid (dashed) line refers to the variance of the fluctuations, as predicted for species $\phi$ ($\psi$). The symbols stand for the homologues quantities as computed via direct stochastic simulations. The Brussellator model is assumed as the reference reaction scheme and parameters refer to the blue square displayed in Fig. \ref{fig2}.}
\label{fig5}
\end{figure}

The linear Langevin equations  (\ref{eq: Langevin_LIN}) are equivalent to the following Fokker-Planck equation for the distribution function $\Pi$ of the fluctuations
\begin{equation}\label{eq:eq Fokker Planck linearizzata vari nodi}
\frac{\partial}{\partial \tau}\Pi =-\sum_{i=1}^{2\Omega}\frac{\partial}{\partial \zeta_i}\left(\mathcal{J}\boldsymbol{\zeta}\right)_i\Pi+\frac{1}{2}\sigma^2\frac{\partial^2}{\partial \zeta_i^2}\Pi
\end{equation}
The solution of the Fokker-Planck equation  is a multivariate Gaussian that we can univocally characterize in terms of the associated first and second moments. It is immediate to show that the first moment converges in time to zero. We focus instead on the the $2 \Omega \times 2 \Omega$ family of second moments, defined as $\langle \zeta_l \zeta_m \rangle = \int \zeta_l \zeta_m \Pi d {\boldsymbol \zeta}$. A straightforward calculation returns:
\begin{widetext}
\begin{equation}
\label{eqn: eq momenti diagonali}
\frac{d}{d\tau}<\zeta_l^2>=2<\left(\mathcal{J}\boldsymbol{\zeta}\right)_l\zeta_l>+\sigma^2=2\sum_{j=1}^{2\Omega}\mathcal{J}_{lj}<\zeta_l\zeta_j>+\sigma^2
\end{equation}
\begin{equation}
\label{eqn: eq momenti misti}
\frac{d}{d\tau}<\zeta_l\zeta_m>=<\left(\mathcal{J}\boldsymbol{\zeta}\right)_l\zeta_m>+<\left(\mathcal{J}\boldsymbol{\zeta}\right)_m\zeta_l>=\sum_{j=1}^{2\Omega}\mathcal{J}_{lj}<\zeta_m\zeta_j>+\mathcal{J}_{mj}<\zeta_l\zeta_j>
\end{equation}
\end{widetext}
for respectively the diagonal and off-diagonal ($l \ne m$) moments. The stationary values of the moments can be analytically computed by setting to zero the time derivatives on the left hand side of equations \eqref{eqn: eq momenti diagonali}-\eqref{eqn: eq momenti misti} and solving the linear system that is consequently obtained. Particularly relevant for our purposes is the quantity  $\delta_i=\langle \zeta_i^2 \rangle$, the variance of the fluctuations displayed, around the deterministic equilibrium, on node $i$. The value of  $\delta_i$, normalized to $\delta_1$, is plotted in Fig. \ref{fig5}, for both species of the Brusselator model. The parameters correspond to the working point identified by the blue square in Fig.  \ref{fig2}. The solid line stands for the, analytically determined, variance of the fluctuations predicted for species $\phi$, while the dashed line refers to species  $\psi$. The symbols are obtained after direct stochastic simulations of system \eqref{eqn: eq di partenza1}-\eqref{eqn: eq di partenza2}, assuming the Brussellator as the reference reaction scheme. As it can be appreciated by visual inspection of  Fig. \ref{fig5}, the fluctuations grow progressively node after node. The predicted variances nicely agree with the result of the simulations on the first nodes of the collection. For $\Omega>15$, deviations are abruptly found and the linear noise approximation fails. Our interpretation goes as follows: the amplification mechanism is manifestly triggered by the imposed noise, which resonates with the peculiar topology of the embedding support. Due to this interplay, noise seeded fluctuations grow across the chain and make it possible for the system to explore the phase space landscape, beyond the local basin of attraction to which it is deterministically bound. 

\begin{figure}
\includegraphics[scale=0.2]{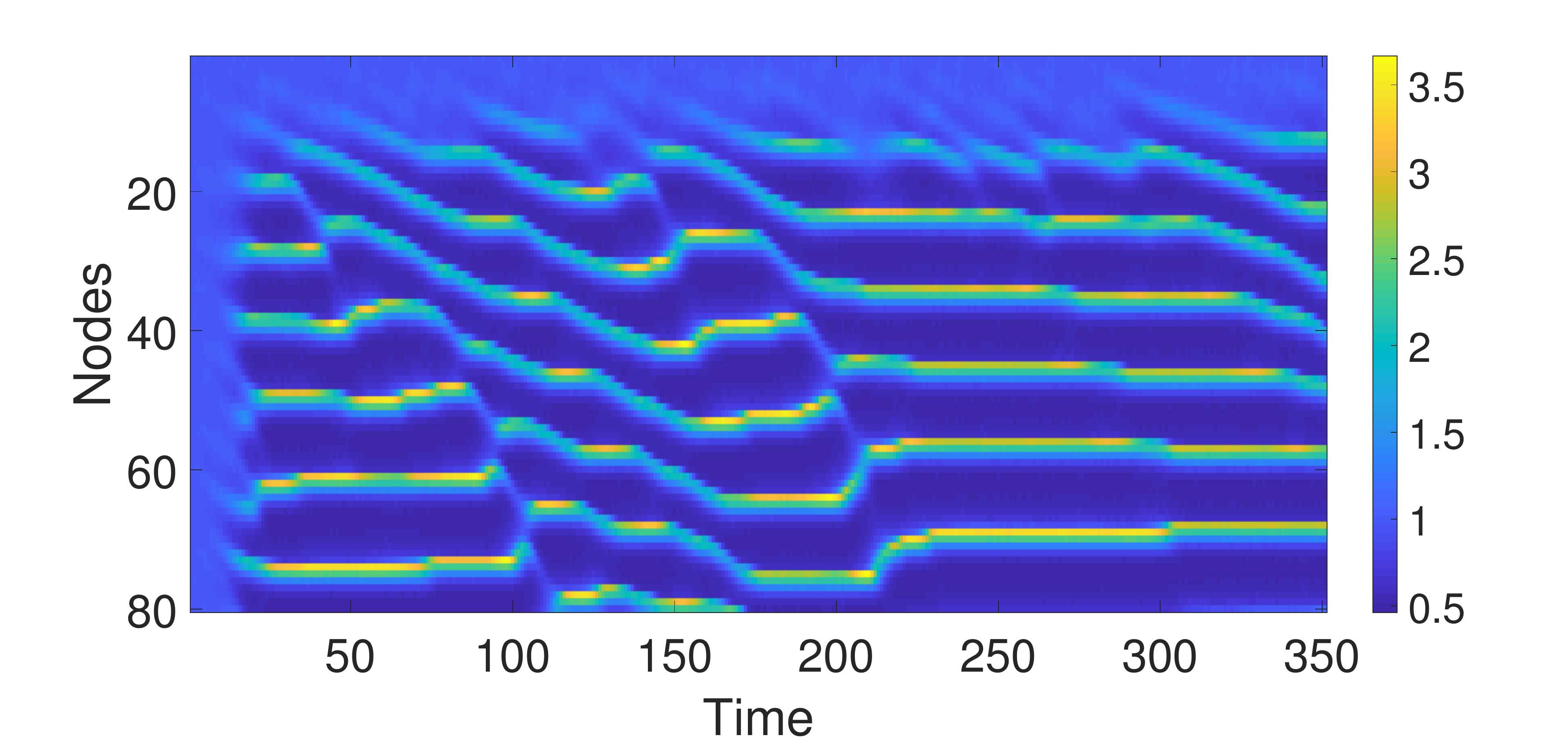} \\
(a) \\
\includegraphics[scale=0.2]{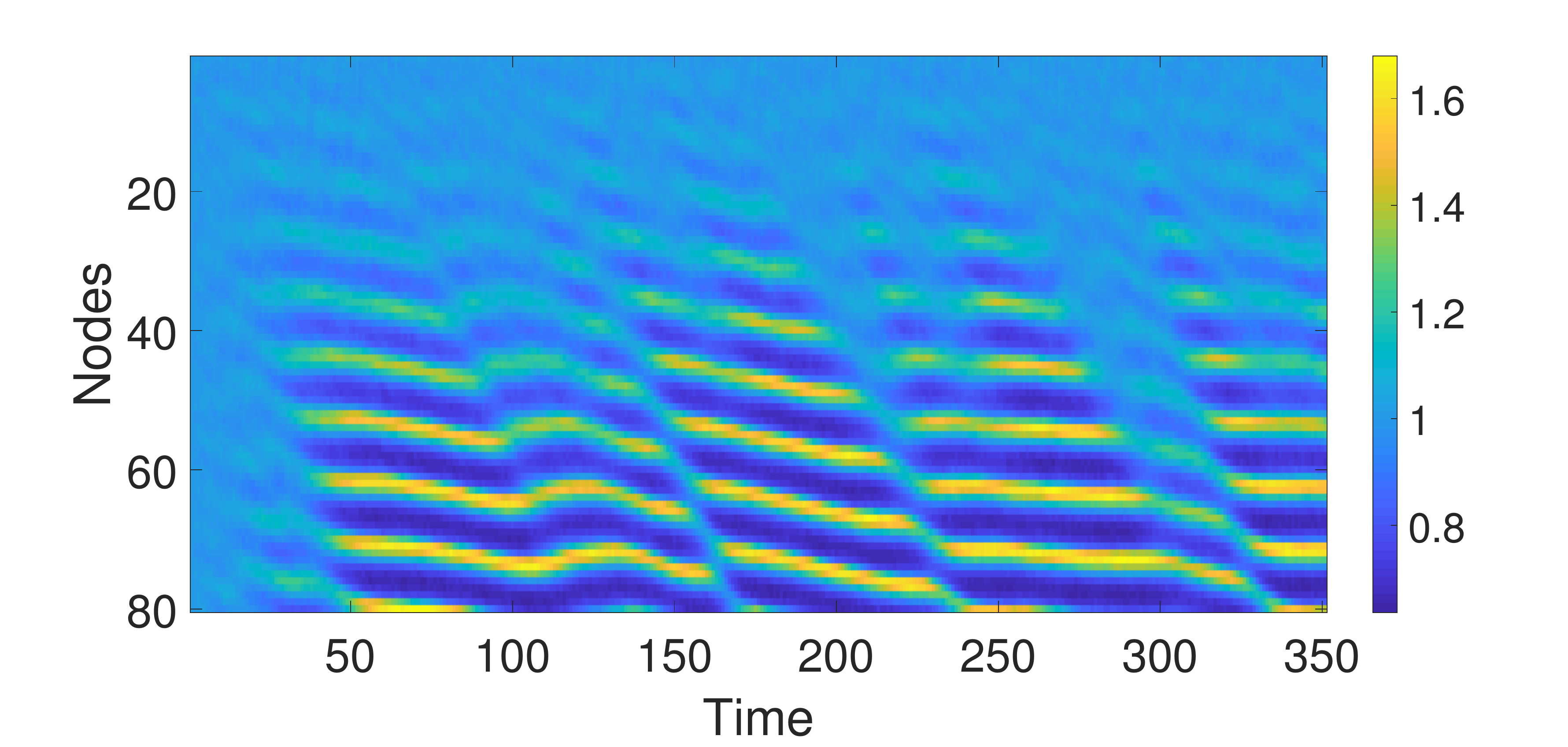} \\
( b)
\caption{The time evolution of species $\phi$ is displayed with an appropriate color code, on different nodes of the chain and against time. The Brusselator reaction scheme is assumed with the same operating choice of Fig. \ref{fig5}. By integrating the stochastic dynamics on a sufficiently long chain yields a robust pattern, which holds permanently, at variance with its deterministic analogue.} 
\label{fig6}
\end{figure}

\begin{figure*}
\begin{tabular}{cc}
\includegraphics[scale=0.4]{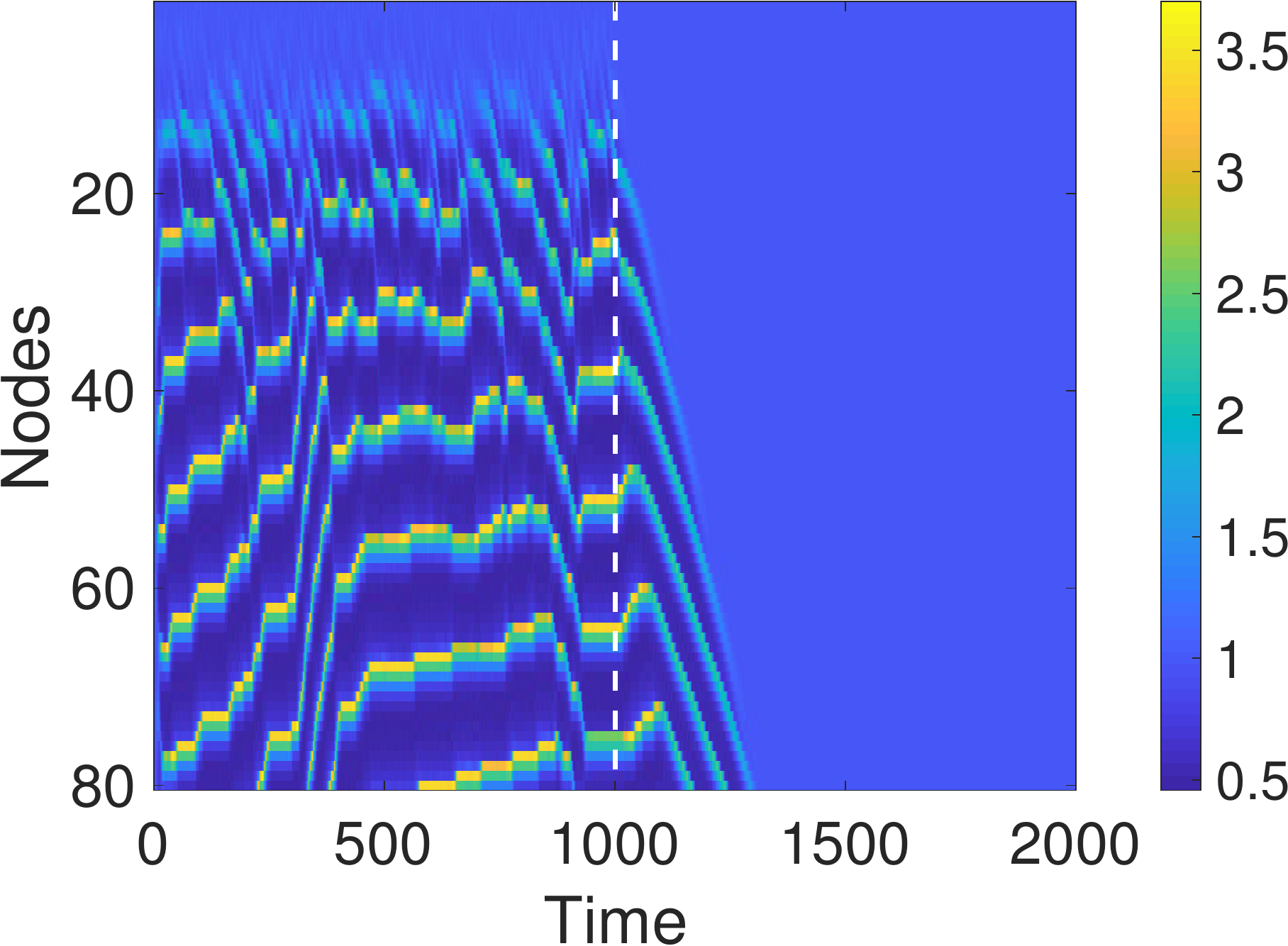} & \includegraphics[scale=0.4]{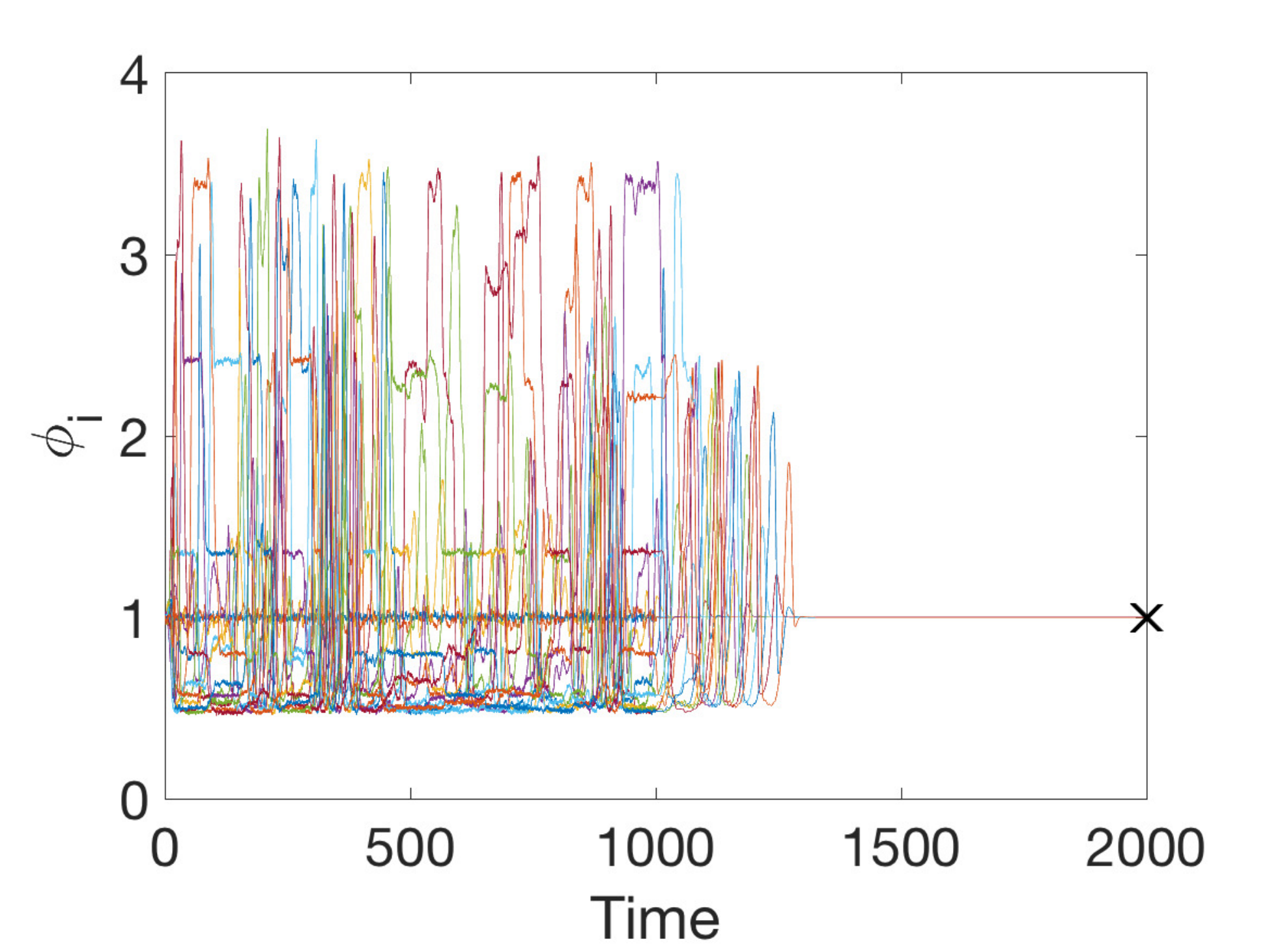} \\
(a) & (b) \\
\includegraphics[scale=0.4]{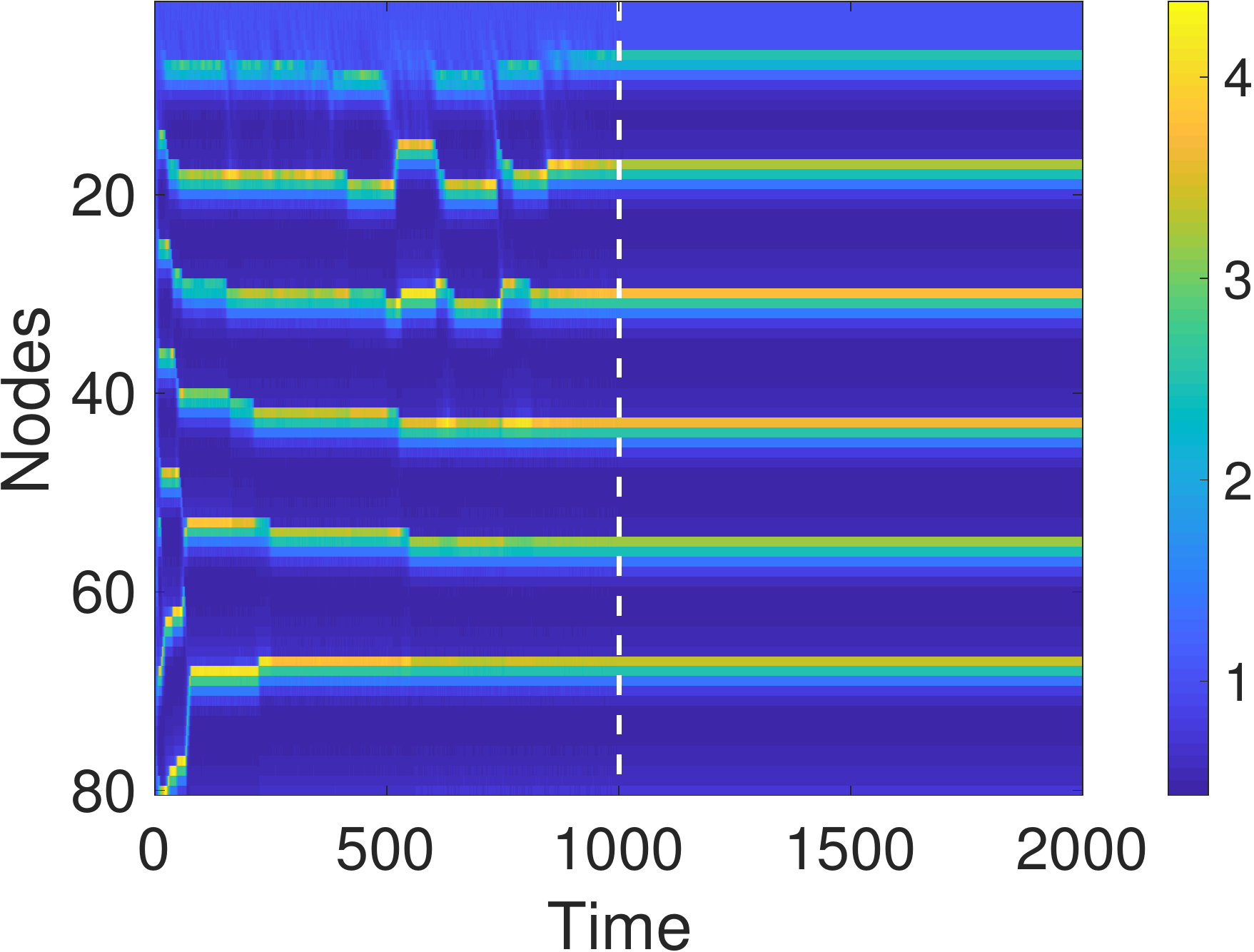} & \includegraphics[scale=0.4]{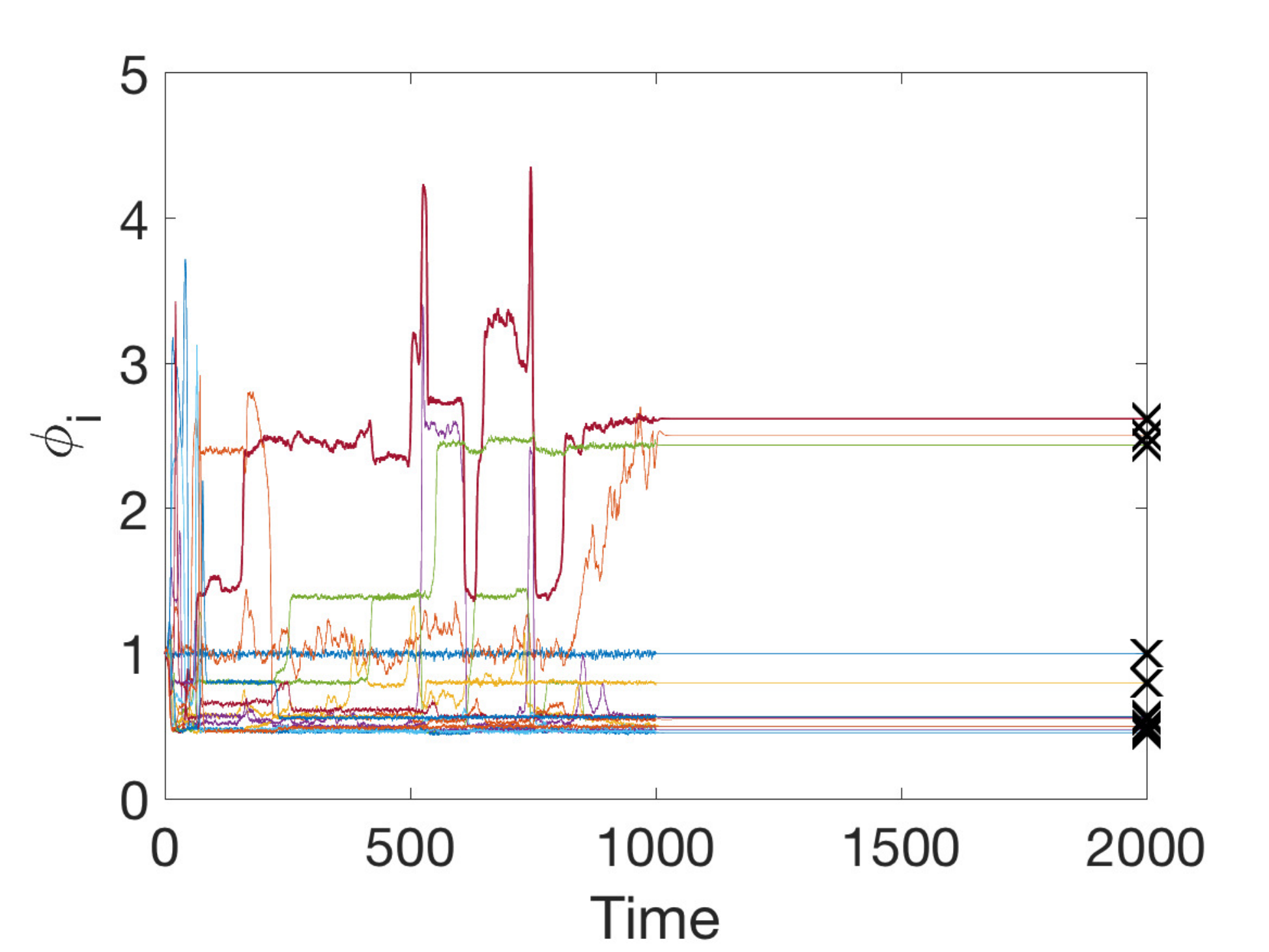} \\
(c) & (d) \\
\end{tabular}
\caption{Panels (a) and (d). The time evolution of species $\phi$ is displayed with an appropriate color code, on different nodes of the chain and against time.  The vertical dashed line identifies the instant in time  when the  external noise is turned off: from here on the system evolves according to a purely deterministic scheme. The Brusselator reaction scheme is assumed with the same parameters choice of Fig. \ref{fig5}, except for $c$ In (a), the pattern fades eventually away. Here $c=2.8$, as in Fig. \ref{fig5}. In (c), the system sediments in a stationary pattern of the Chimera type. Here $c=2.4$. In Figs. (b) and (c),
the density of species $\phi$  is displayed on a few nodes of the collection (one node each five, across the chain). In panel (b) (corresponding to pattern (a)), the system converges to the homogeneous fixed point (black cross). In panel (d) (corresponding to pattern (b)), the system reaches a stable heterogeneous attractor.} 
\label{fig7}
\end{figure*}

From here on,  it is no more legitimate to simplify the dynamics of the system as if it was evolving in the vicinity of the homogeneous solution and the assumption that sits at the root of the linear noise estimate are consequently invalidated.  The time evolution of species $\phi$ is plotted, with an appropriate color code, on different nodes of the chain and versus time in  Fig. \ref{fig6}. Noise secures the stabilization of complex dynamical patterns, which are perpetually maintained in the stochastic version of the model, so breaking the spatial symmetry that characterizes the asymptotic deterministic solution. The amplification mechanism driven by the stochastic component of the dynamics can solely occur within a closed domain of the parameter space $(b,c)$ adjacent to the region of deterministic instability. The domain of interest is  delimited  by the red solid line in Fig. \ref{fig2}: for the parameters that fall below the red line, the variance of the fluctuations is analytically predicted to increase along the chain.  Even more importantly,  the system may be frozen in the heterogenous state, when silencing the noise (so regaining the deterministic limit) after a transient of the stochastic dynamics. This is clearly the case provided the deterministic dynamics possesses a stable non homogeneous attractor, for the chosen parameters set. In Fig.  \ref{fig7} (a) the pattern is stably displayed, only when the noise is active. When the stochastic forcing is turned off (at the time identified by the white dashed line),  the system regains the homogenous equilibrium. A selection of individual trajectories, recorded on specific nodes of the chain, is plotted in Fig. \ref{fig7} (b) and yields the same qualitative conclusion. A different scenario is instead met when operating with a slightly smaller value of the parameter $c$ (still outside the region of deterministic Turing instability). When turning off the noise, the system spontaneously sediments in a Chimera like pattern, see Fig. \ref{fig7} (c),  a superposition of homogeneous (in the beginning of the chain) and heterogenous states (at the bottom of the chain) \cite{chimera1,chimera2,chimera3}. The same conclusion is reached upon inspection of Fig.  \ref{fig7} (d). Here, the deterministic attractor is represented with a collection of crosses. A straightforward calculation confirms that it is indeed one of the different non homogeneous and stable attractors displayed by the system in its deterministic version.

\section{Quasi-degenerate directed lattice}

This section is devoted to generalize the above analysis to the relevant setting where the degeneracy of the problem is removed by the insertion of return links, among adjacent nodes. More specifically we will assume that an edge with weight $\epsilon$ exists that goes from node $i$ to nodes $i-1$, for all $i>1$. At the same time the strength of the corresponding forward link is set to $1-\epsilon$, so as to preserve the nodes' strength (for $i>1$), when modulating $\epsilon$. In the following we will consider $\epsilon$ to be small. In particular, for $\epsilon \rightarrow 0$ one recovers the limiting case discussed in the previous section. On the other hand the introduction of a tiny return probability suffices to break the degeneracy of the problem: the $\Omega$ eigenvalues become distinct and the eigenvectors of the Laplacian define a basis 
which can be used to solve the linear problem (\ref{linear_syst}), which stems for the deterministic version of the inspected model. Denote by $\mathbf{v^{\alpha}}$ the eigenvector of $\mathbf{\Delta}$ relative to eigenvalue $\Lambda^{(\alpha)}$, with $\alpha=1,.., \Omega$. In formulae,   $\sum_{j=1}^{\Omega}\Delta_{ij}v_j^{\alpha}=\Lambda^{\alpha}v_i^{\alpha}$. The perturbation  $\boldsymbol{\zeta}$ in equation (\ref{linear_syst}) 
can be expanded as $\zeta_i=\sum_{\alpha=1}^{\Omega} c_{\alpha} e^{\lambda_{\alpha}t}v_i^{\alpha}$, where the constants $c_{\alpha}$ are determined by the initial condition. By inserting the aforementioned ansatz into the linear system (\ref{linear_syst}) yields the  self-consistent condition 
\begin{equation}
    \det
    \begin{pmatrix}
    f_{\phi}+D_{\phi}\Lambda^{\alpha}-\lambda_{\alpha} & f_{\psi} \\
    g_{\phi} & g_{\psi}+D_{\psi}\Lambda^{\alpha}-\lambda_{\alpha}
    \end{pmatrix}
=0.
\end{equation}

\begin{figure}
\includegraphics[scale=0.2]{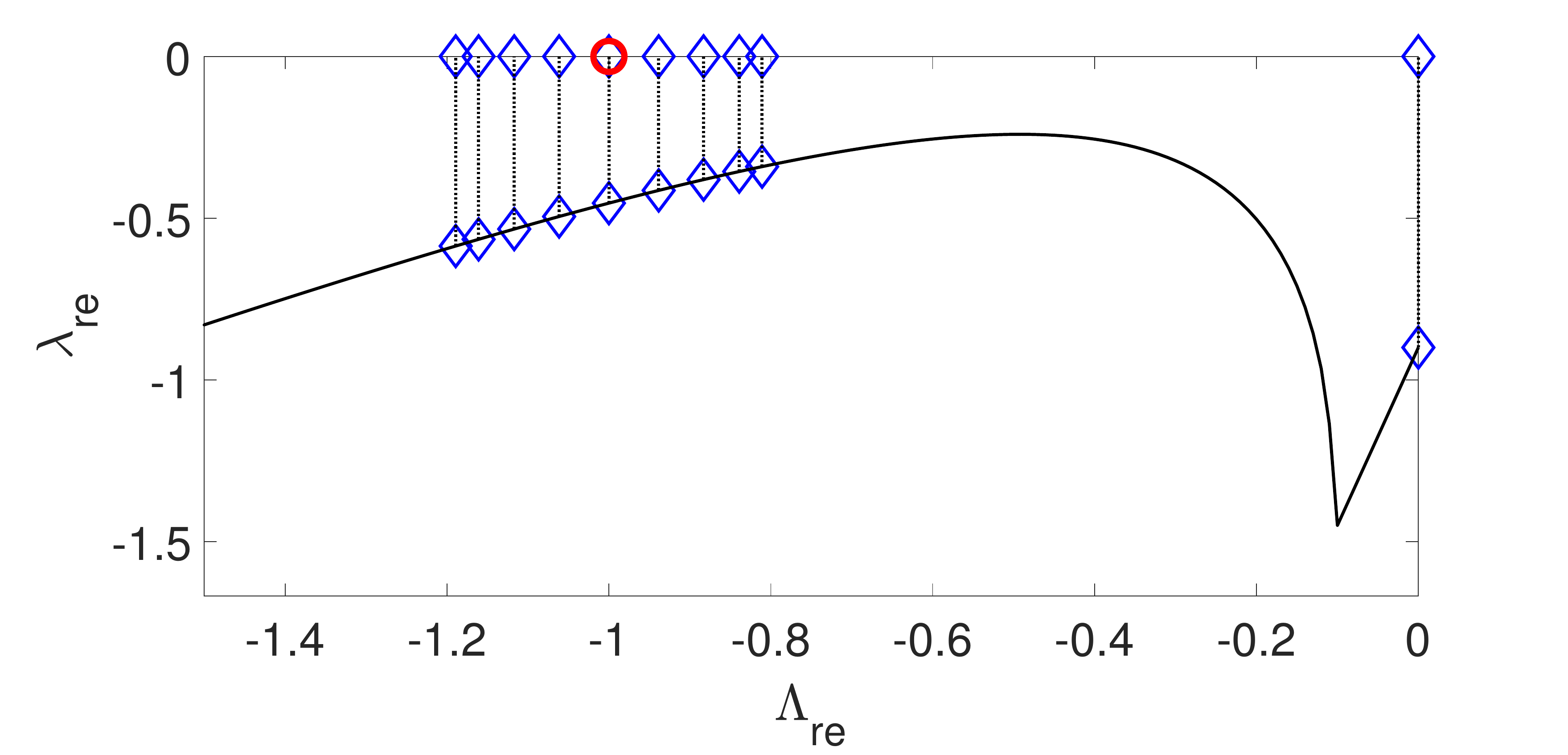}\\
(a)\\
\includegraphics[scale=0.2]{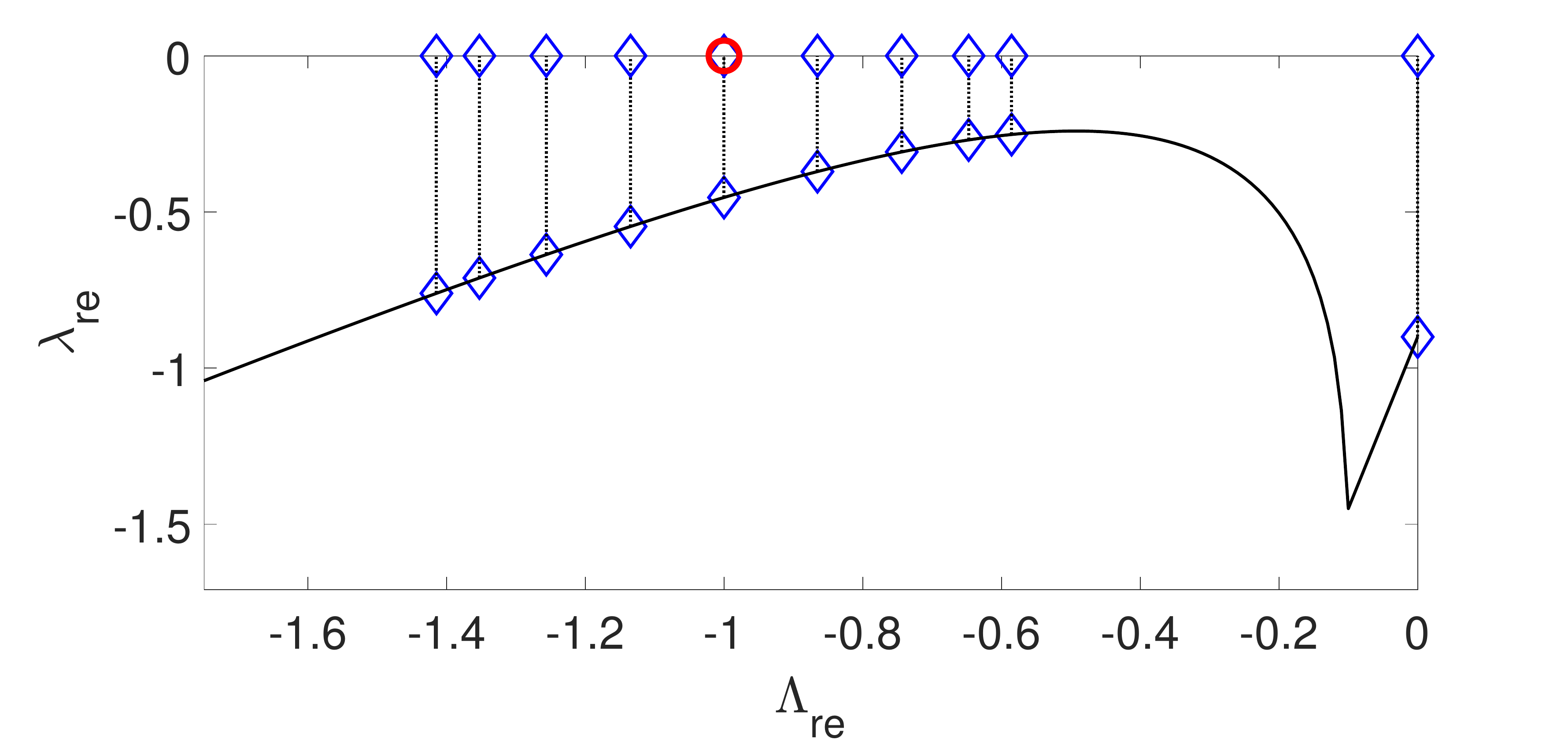}\\
(b)\\
\includegraphics[scale=0.2]{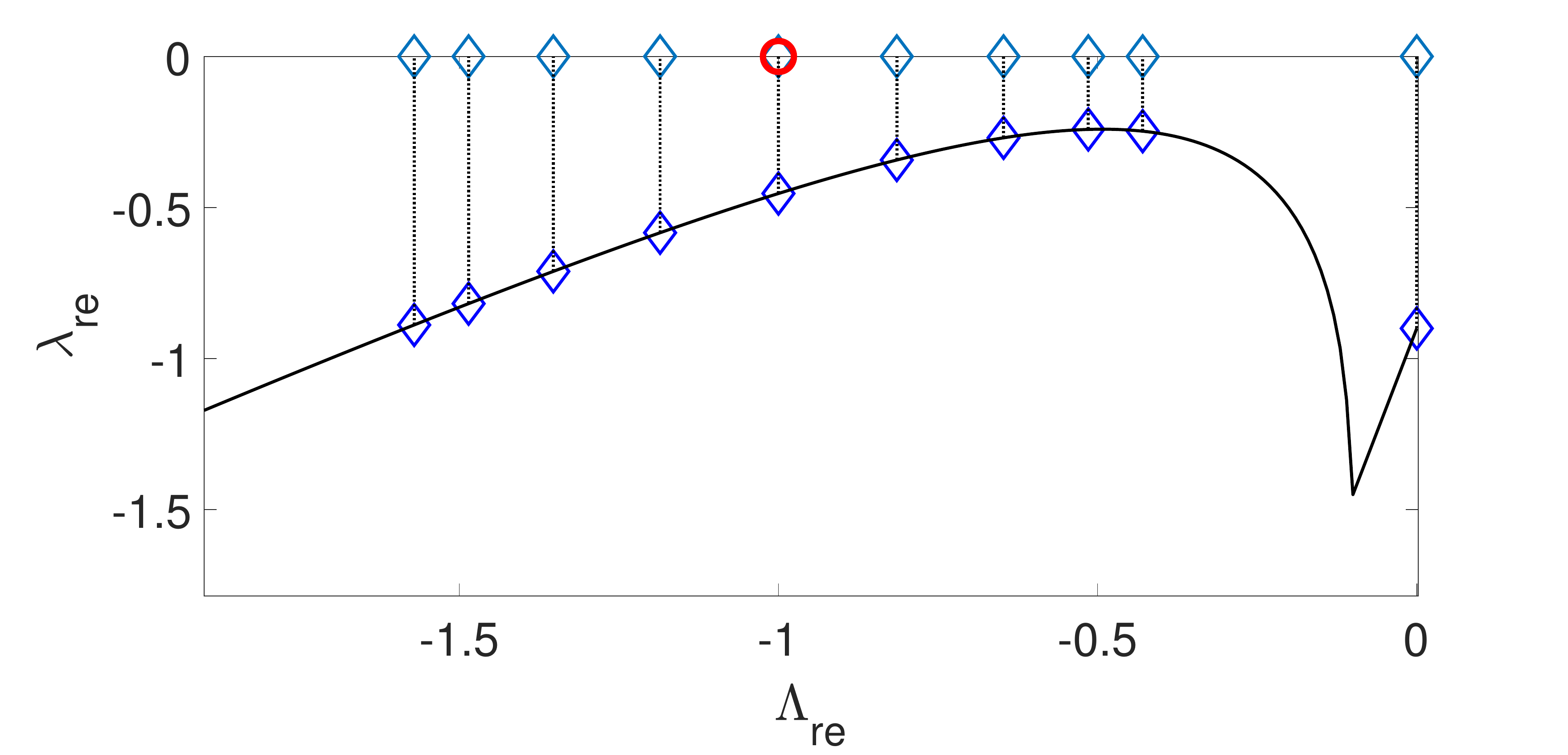}\\
(c)
\caption{Laplacian eigenvalues on the real axis in correspondence with their respective points on the dispersion relation for different values of $\epsilon$ [(a) $\epsilon=0.01$, (b) $\epsilon=0.05$, (c) $\epsilon=0.1$] and $\Omega=10$. 
The solid lines stand for the dispersion curve obtained when placing the system on a continuous spatial support. Vertical dashed lines are a guide for the eye to project the discrete dispersion relation back to the real axis where the spectrum of the Laplacian falls. The red circle is position at $-1$ the vallue to which the eigenvalues tend when sending $\epsilon \rightarrow 0$.
}
\label{fig8}
\end{figure}
The stability of the homogenous fixed point can be therefore determined from the above condition, by computing $\lambda_{\alpha} $  as a function of the Laplacian eigenvalues $\Lambda^{(\alpha)}$. This is the generalization of the so called dispersion relation to a setting where the spatial support is a network. In Fig. \ref{fig8}, we plot the dispersion relation for a chain made of $\Omega=10$ nodes and for different values of $\epsilon$, assuming the Brusselator model as the reference scheme. The reaction parameters are set so as to yield the square symbol in Fig.\ref{fig2}. Remarkably, the spectrum of the Laplacian operator is real and the largest eigenvalue is $\Lambda^{(1)}$=0, as it readily follows by its definition.  The remaining $\Omega-1$ eigenvalues are real and cluster in the vicinity of $\bar{\Lambda}=-1$, the smaller $\epsilon$ is, the closer they get to this value, as illustrated Fig \ref{fig8}. The case of a degenerate chain can be formally recovered by sending $\epsilon$ to zero, which in turn implies that the non trivial portion of the dispersion curve, as depicted in Fig \ref{fig8}, collapses towards an asymptotic attractor located at $(\bar{\Lambda},\lambda(\bar{\Lambda}))$. Building on this observation, it can be shown that the solution of the deterministic linear problem (\ref{linear_syst}), for the system defined on a degenerate chain, can be obtained by performing the limit  for $\epsilon \rightarrow 0$ of the non degenerate linear solution.

It is hence tempting to speculate that the stochastic driven instability as outlined in the preceding section, can readily extend to a setting where the chain is non degenerate, provided $\epsilon$ is sufficiently small. The remaining part of this section is entirely devoted to explore this interesting generalization . 

\begin{figure}
\includegraphics[scale=0.2]{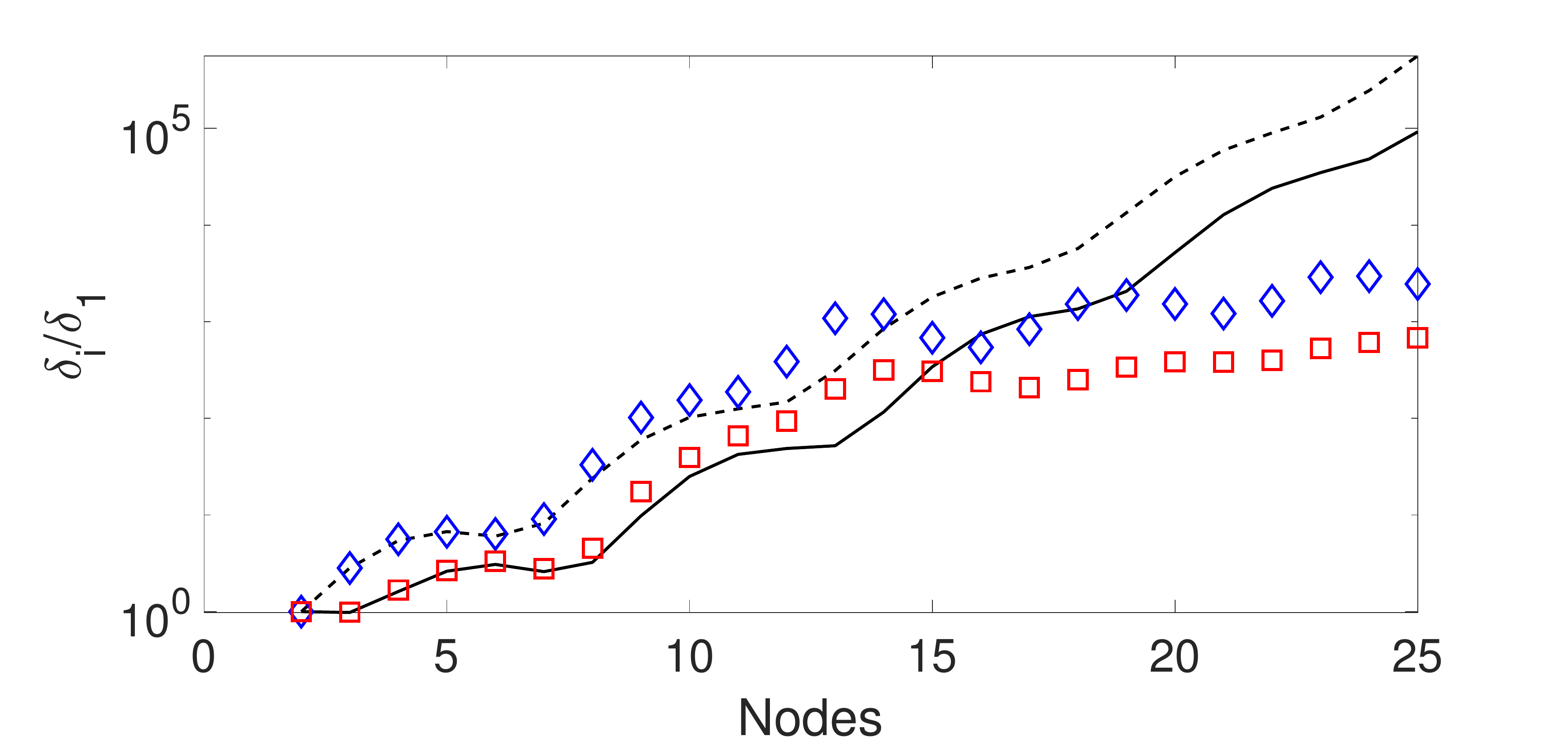}\\
\caption{$\delta_i/\delta_1$ is plotted against the node's number along the chain. Here we set $\epsilon=0.01$. The solid (dashed) line refers to the variance of the fluctuations as predicted for species $\phi$ ($\psi$). The symbols stand for the homologues quantities as computed via direct stochastic simulations. The parameters refer to the  blue square displayed in Fig. \ref{fig2}.}
\label{fig9}
\end{figure}

\begin{figure}
\includegraphics[scale=0.2]{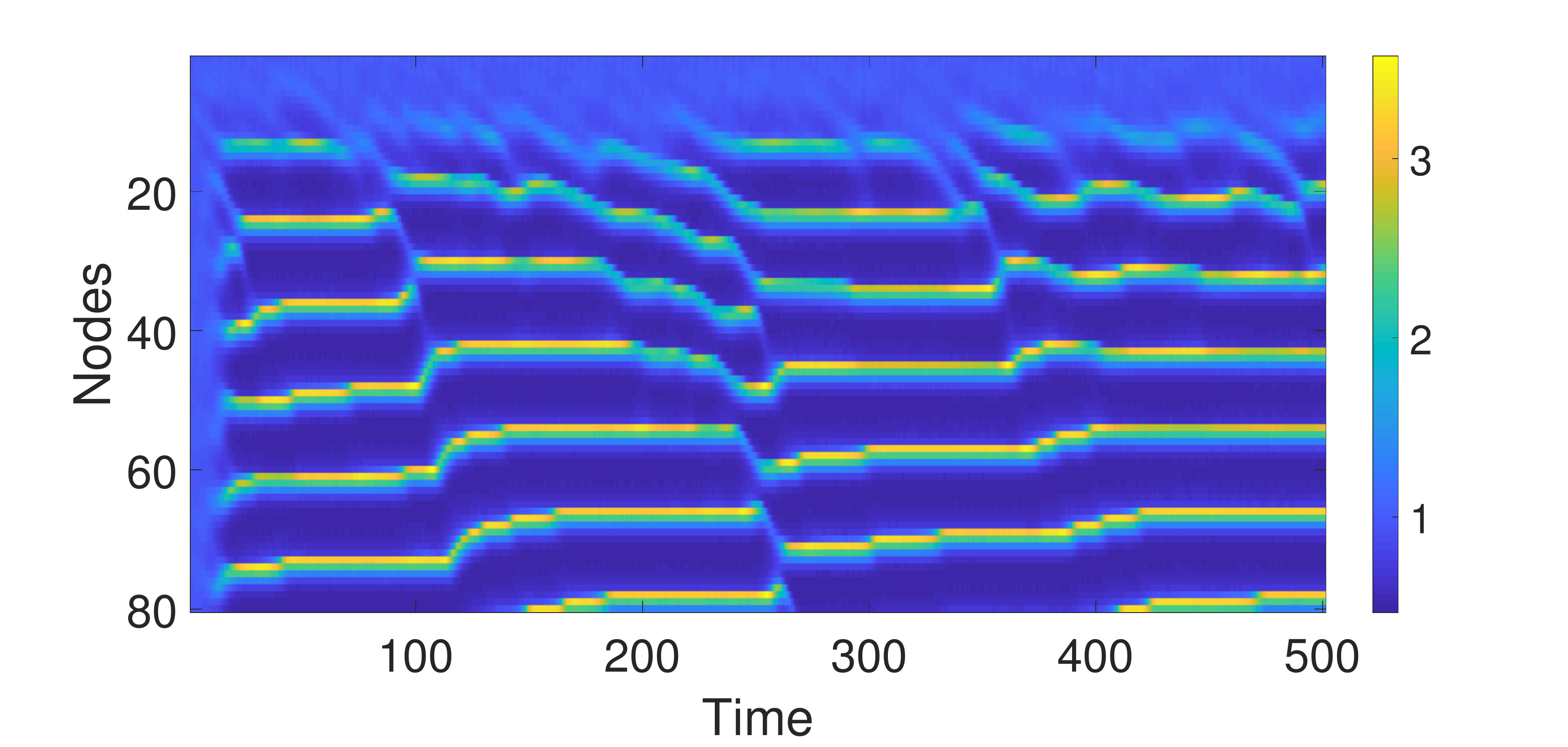}
\caption{The time evolution of species $\phi$ is displayed with an appropriate color code, on different nodes of the chain and against time. Here $\epsilon=0.01$ and the Brusselator reaction scheme is assumed with the same parameters choice of Fig. \ref{fig9}}
\label{fig10}
\end{figure}

Following the strategy discussed above, we can set to calculate the stationary values for the moments of the stochastic fluctuations.  In Fig. \ref{fig9}, the stationary values of the moments $\delta_i=\langle \zeta_i^2 \rangle$ are normalized to $\delta_1$ and plotted against the node label across the chain. In analogy with the above, the solid line stands for the variance of the fluctuations associated to species $\phi$, while the dashed line stands for species  $\psi$. As expected, the fluctuations magnify along the chain, as it happens when the system is made to evolve with $\epsilon=0$. The predicted variances agree with the results of the stochastic simulations, up to a critical length of the chain above which the system begins feeling the non linearities that will steer it towards a non homogeneous attractor. Noise stabilizes the heterogeneous patterns which become hence perpetual in the stochastic version of the model, as it can be appreciated by inspection of Fig. \ref{fig10}.

\begin{figure}
\includegraphics[scale=0.2]{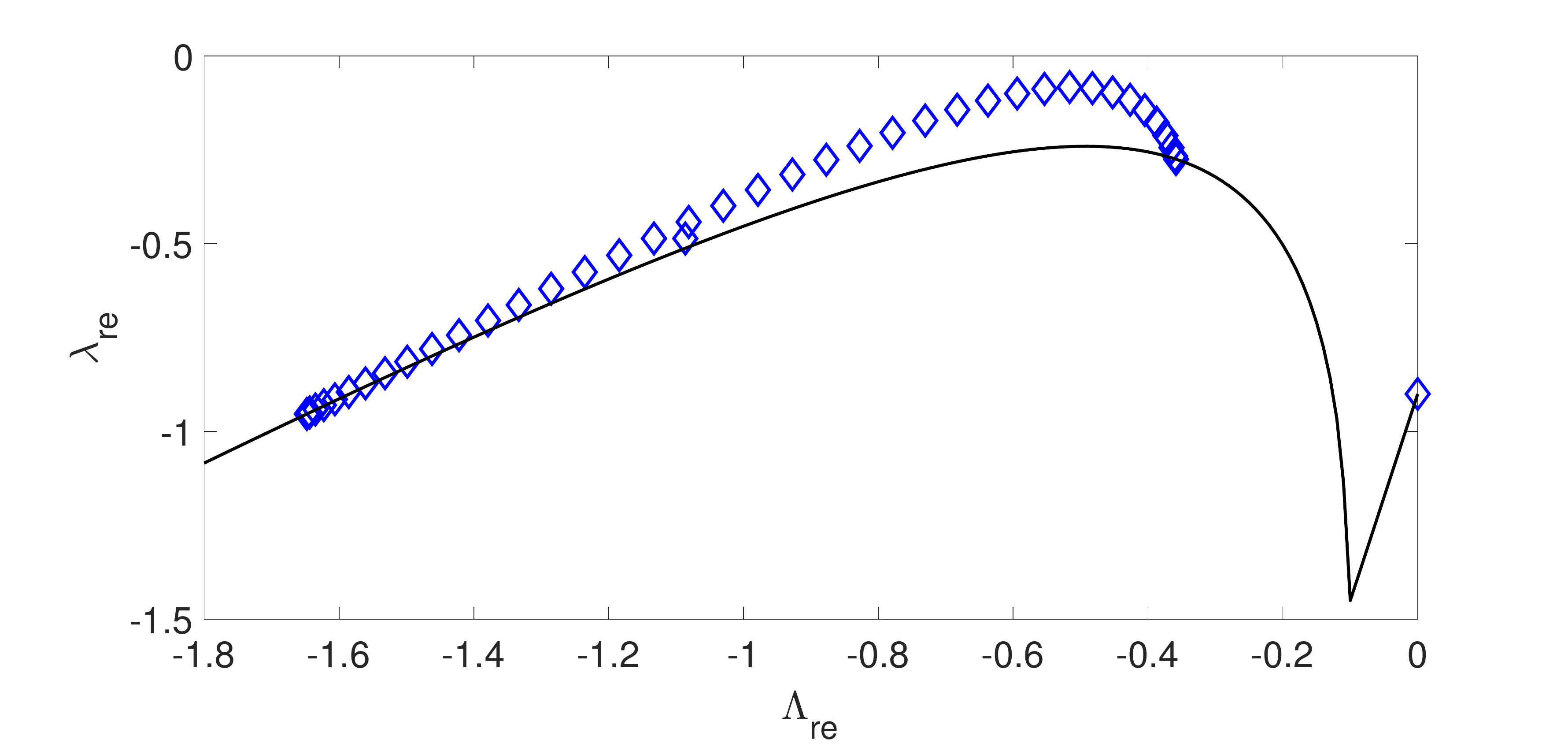}
\caption{Dispersion relation for $\epsilon$ selected randomly, for each couple of nodes, from a Gaussian distribution with mean $\bar{\epsilon}=0.01$ and variance $\eta_{\bar{\epsilon}}=10^{-4}$. Here the chain is assumed to be $\Omega=80$ nodes long.}
\label{fig11}
\end{figure}

\begin{figure}
\includegraphics[scale=0.2]{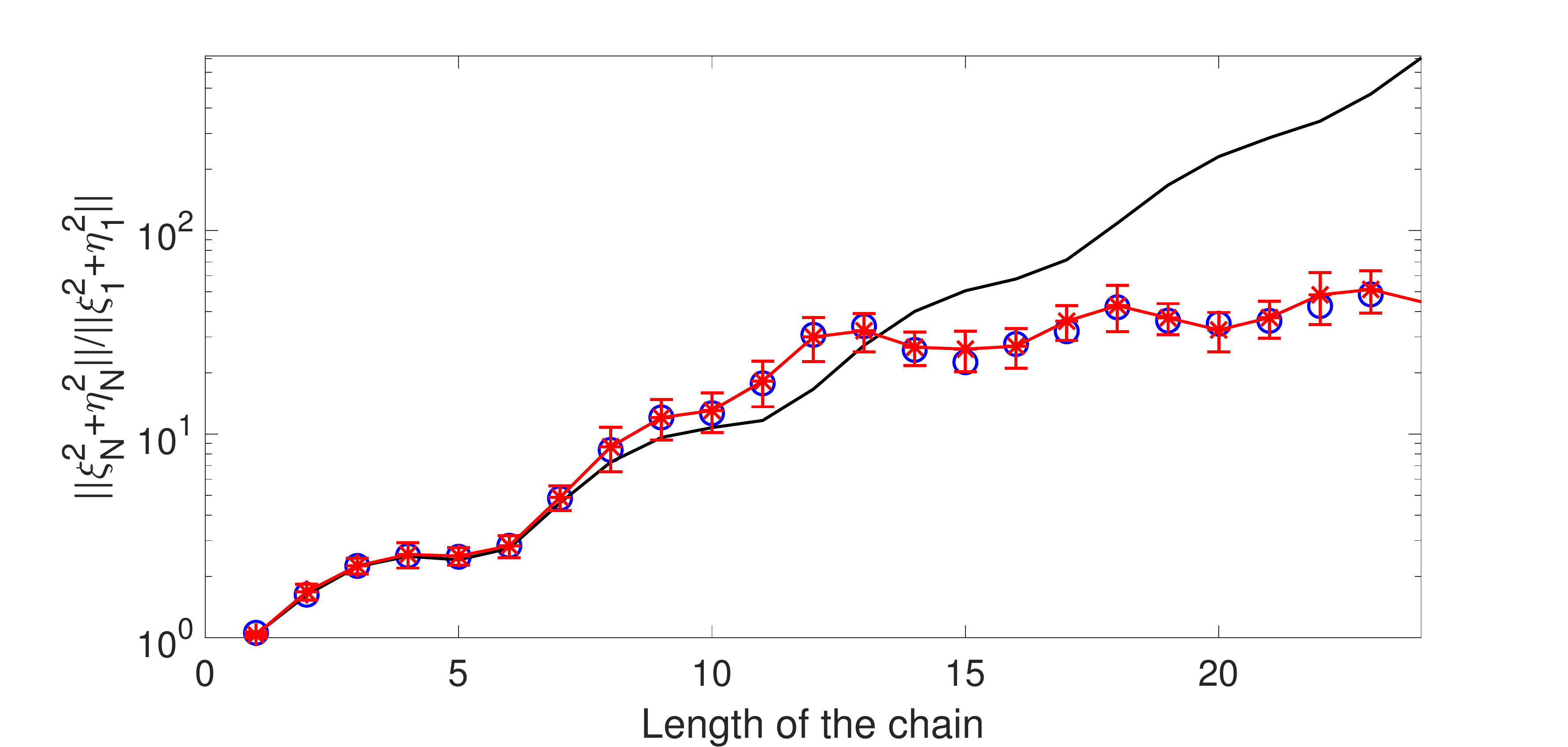}
\caption{The ratio between the norm of the fluctuations on the last and first nodes of the chain is plotted against the length of the chain. The solid line refers to the norm as predicted. The symbols stand for the homologous quantities as computed via direct stochastic simulations. Blue diamonds refer to fixed weights $\epsilon=0.01$, red squares to random weights (averaging over $25$ realizations) chosen from a Gaussian distribution centered in $\bar{\epsilon}=0.01$, with variance $\eta_{\bar{\epsilon}}=10^{-4}$. The parameters are the same as those in Fig. \ref{fig9}.}
\label{fig13}
\end{figure}

\begin{figure}
\includegraphics[scale=0.2]{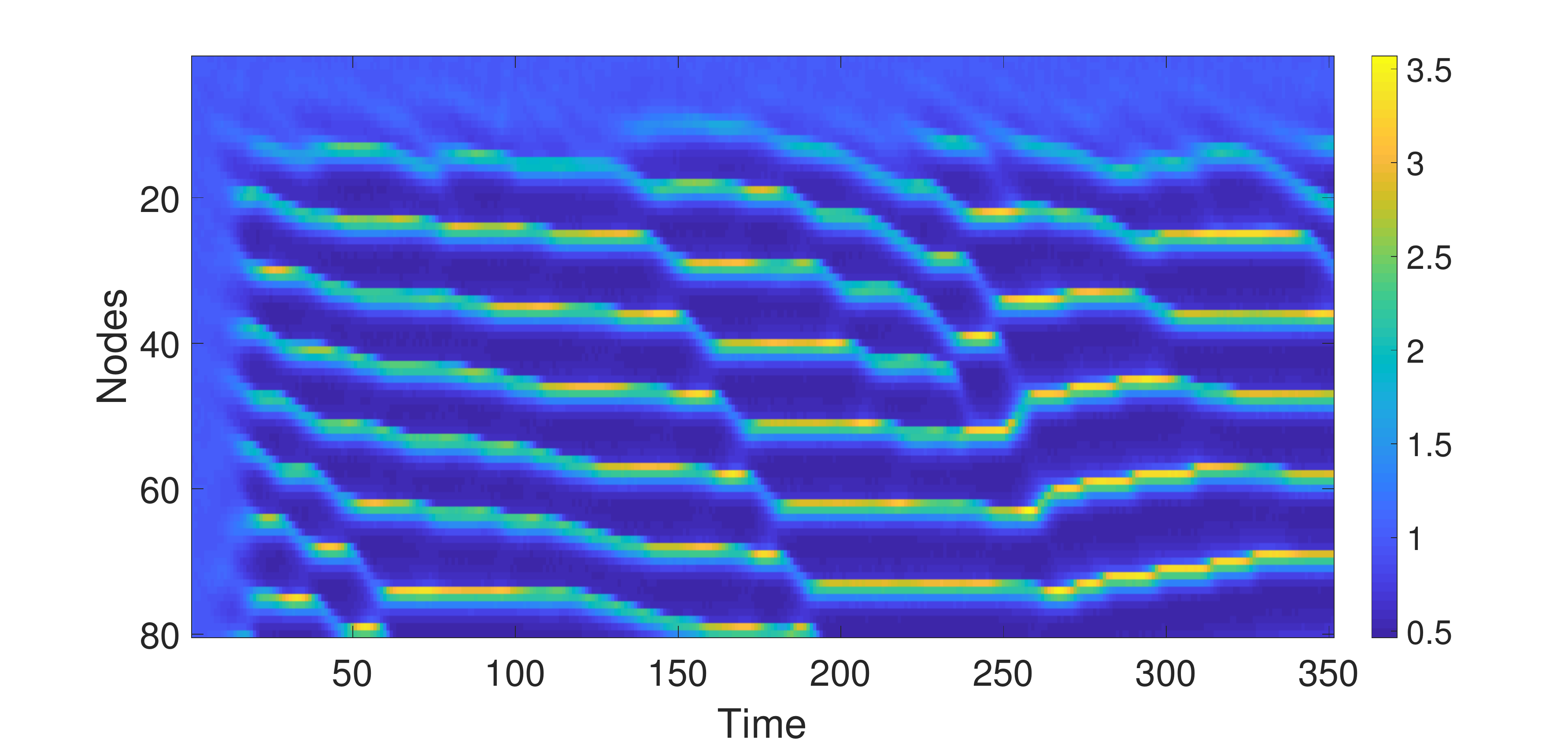}
\caption{The time evolution of species $\phi$ is displayed with an appropriate color code, on different nodes of the chain and against time. Here we assume the weights $\epsilon$ to be random entries chosen from a Gaussian distribution centered in $\bar{\epsilon}=0.01$ with variance $\eta_{\bar{\epsilon}}=10^{-4}$. The parameters are the same as those in Fig. \ref{fig9}.}
\label{fig12}
\end{figure}

As a further attempt to grasp the complexity of the phenomenon, we consider again a chain with return links, but assume the weights $\epsilon$ to be random entries drawn from a Gaussian distribution, centered in $\bar{\epsilon}$, with variance $\eta_{\bar{\epsilon}}$. The spectrum of the Laplacian operator is now complex (at variance with the case where the $\epsilon$ are uniform) and the discrete dispersion relation is no longer bound to the idealized continuum curve, see Fig. \ref{fig11}. For a sufficiently small  $\bar{\epsilon}$, and modest scattering around the mean, the negative portion of the dispersion relation clusters in the vicinity of the point $(\bar{\Lambda},\lambda(\bar{\Lambda}))$, which stems from the fully degenerate solution. Arguing as above, one can expect that noise and spatial coupling will cooperate also in this setting to yield robust stochastic patterns, in a region of the parameters for which deterministic stability is granted. In Fig. \ref{fig13}, we show that the signal amplification for the stochastic Brusselator model defined on a chain with random return weights, is on average identical to that observed when the $\epsilon$ are uniform and set to the average value $\bar{\epsilon}$. The depicted points are computed after averaging over $100$ realization of the random quasi degenerate network and the error in the computed quantities is of the order of the symbols size.  The ensuing stochastic pattern is reported in Fig. \ref{fig12}.

\begin{figure}
\includegraphics[scale=0.2]{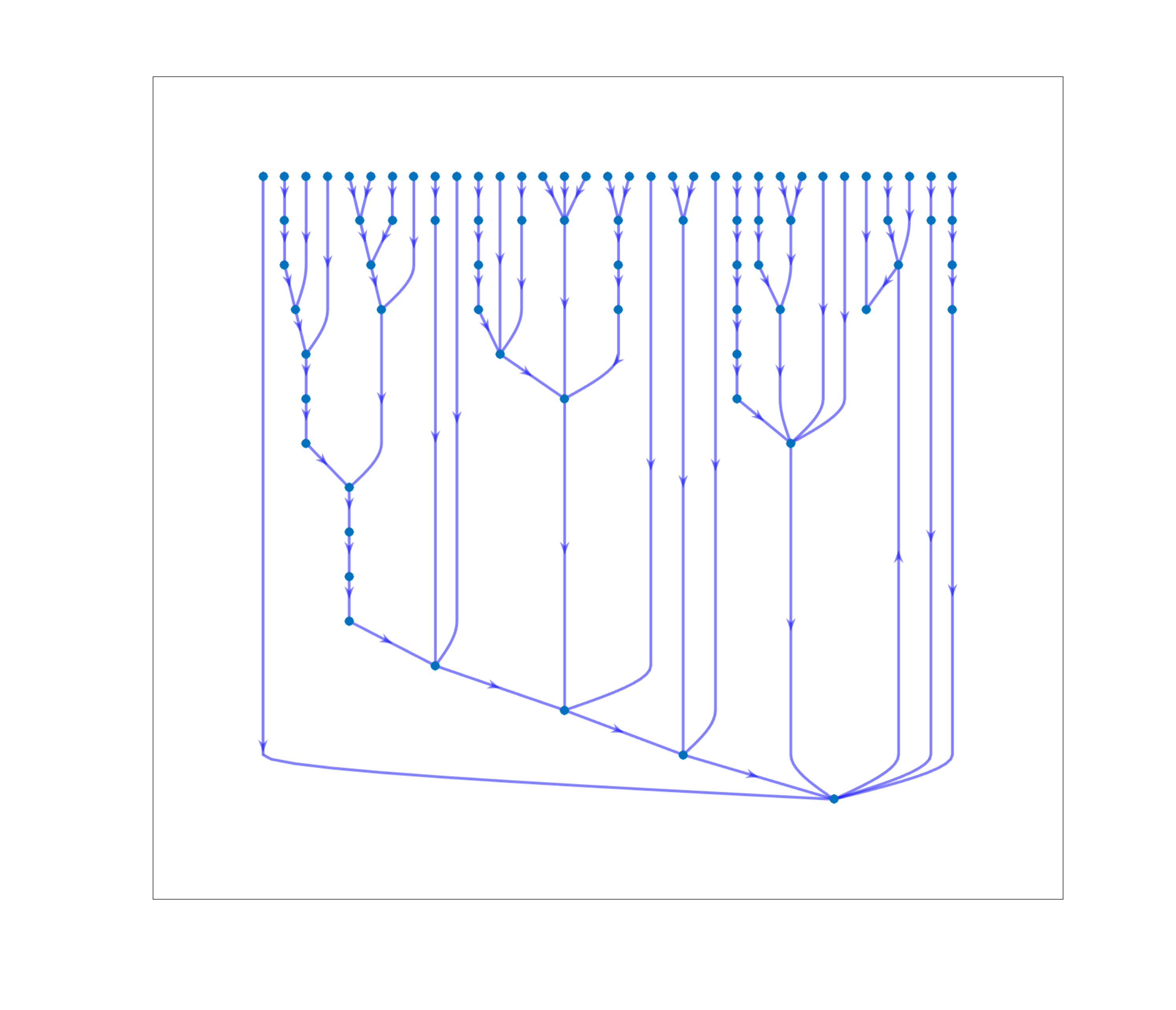}
\caption{Example of network generated with the procedure described in the text made of $\Omega=80$ nodes. The Brusselator reaction scheme is assumed with the parameter choice of blue square in Fig. \ref{fig2}.}
\label{fig14}
\end{figure}

\begin{figure}
\includegraphics[scale=0.2]{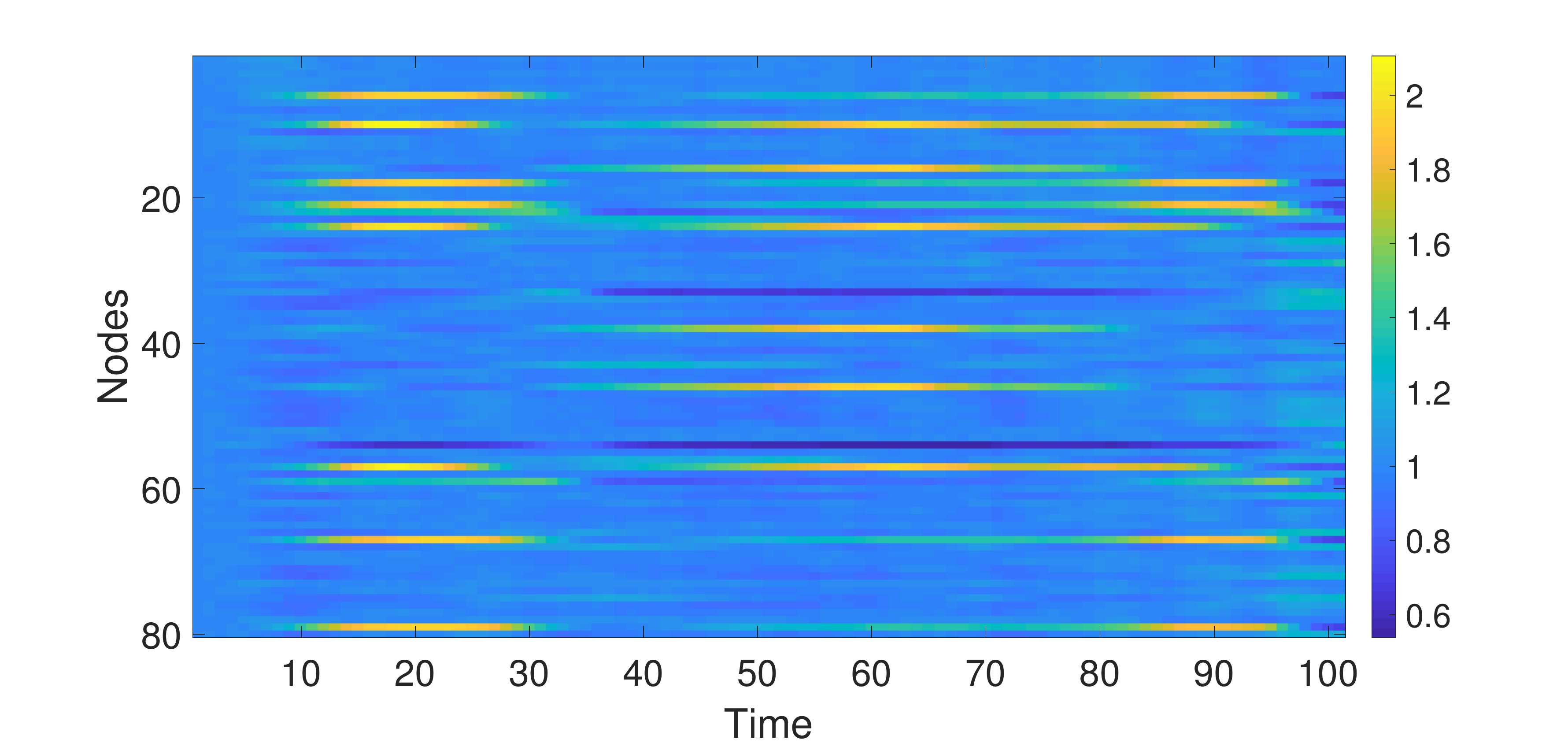}
\caption{The time evolution of species $\phi$ is displayed with an appropriate color code, on different nodes of the network of Fig. \ref{fig14} and against time.}
\label{fig15}
\end{figure}

In conclusion, we have shown that a generic reaction model, which is stable when defined on a continuous or lattice-like support, can turn unstable due to the cooperative interplay of two effects, noise and the 
quasi degenerate nature of generalized Jacobian operator, as reflecting the specific spatial support here assumed. An increase in the node number yields a progressive amplification of the fluctuations on the rightmost end of the direct chain, a process which eventually drives the uniform attractor unstable. In the reciprocal space, one gains a complementary insight into the scrutinized phenomenon. Driving stochastically unstable a system, which is deemed stable under the deterministic angle, requires packing within a bound domain of the complex plane a large collection of Laplacian eigenvalues. The eigenmodes associated to the quasi degenerate 
spectrum provide effective route system to vehiculate the instability. One could therefore imagine to generate networks prone to the instability, by hierarchically assembling nodes in such a way that the associated Laplacian possesses a quasi degenerate spectrum, according to the above interpretation. To challenge this view in the simplest scenario possible, we implemented a generative scheme which builds on the following steps. First we consider two nodes, linked via a direct edge that goes from the first to the second. The Laplacian associated to this pair displays two eigenvalues, one in zero and the other localized in $-1$. We then add a third node to the collection.  We select at random a node,  from the pool of existing ones, and identify it as target of a link that originates from the newly added node. The strength of the link is chosen randomly from a Gaussian distribution with given mean and standard deviation. We then compute the spectrum of the obtained network and accept  the move if the new eigenvalue falls sufficiently close to $-1$, 
or reject it otherwise. The procedure is iterated for a maximum number of times which scales extensively with the size of the network. Once the third node is aggregated to the initial pair, we move forward to adding the fourth node according to an identical strategy that we iterate forward. One exemplary of networks generated according to this procedure is displayed in Fig. \ref{fig14}. This is the skeleton of a distorted one dimensional directed chain,  short segments being added on the lateral sides, and fall in the class of the so called random directed acyclic graphs. The Brussellator model evolved on this network (see Fig. \ref{fig15}) returns noise triggered patterns which share similar features with those obtained earlier. In principle, one could devise more complicated networks, that would return a patchy distribution of eigenvalues, engineered to densely populate distinct regions of the complex plane. This extension is left for future work.

\section{Conclusions}
To investigate the stability of an equilibrium solution of a given dynamical system, it is customary to perform a linear stability analysis which aims at characterizing the asymptotic evolution of an imposed perturbation. In doing so, one obtains a Jacobian matrix, evaluated at the fixed point of interest, whose eigenvalues bear information on the stability of the system. If the eigenvalues of the Jacobian matrix display negative real parts, the system is resilient, meaning that it will eventually regain the equilibrium condition, by exponentially damping the initial perturbation. Non-normal reactive components may however yield a short time amplification of the disturbance, before the crossover to the exponential regime that drives the deterministic system back to its stable equilibrium. Particularly interesting is the interplay between noise, assumed as a stochastic perpetual forcing, and the inherent non-normality, as stemming from the existing interactions. Patterns  have been for instance reported to occur for spatially extended stochastic models, with a pronounced degree of non-normal reactivity and outside the region of deterministic instability. Building on these premises,  we have here taken one step forward towards the interesting setting where the degree of inherent non-normality is magnified by the embedding spatial support. Indeed, by replicating a two-species model on diffusively coupled patches of a directed lattice, we enhanced the ability of the system to grow perturbation at short time. This effect is the byproduct of the degeneracy in the spectrum of the Jacobian matrix associated to the examined system, and which ultimately reflects the architecture of couplings between adjacent units.  A non trivial amplification of the noise across the lattice is observed and explained, which materializes in self-organized patterns, that are instead lacking in the deterministic analogue of the analyzed model. Our conclusions are then extended to a quasi-degenerate support: the ingredient that we have identified as crucial for the onset of the amplification is the presence of a compact region in the complex plane where the eigenvalues of the Laplacian operators accumulate. Beyond a critical size of the system, expressed in terms of number of nodes that define the support, the system may lose its deterministic resilience. In fact, it can eventually migrate towards another attractor that is stably maintained, also when the noise forcing is turned off. Taken all together, our investigations point the importance of properly accounting for the unavoidable sources of stochasticity when gauging the resilience of a system: non normality and quasi degenerate networks might alter dramatically the deterministic prediction turning unstable a system that would be presumed otherwise  stable under a conventional deterministic perspective. Furthermore, our findings bear a remarkable similarity with the phenomenon of convective instability, \cite{deissler, lepri1, lepri2, jiotsa}, a possible connection that we aim at investigating in a future contribution.


\begin{thebibliography}{99}
\bibitem{resilience1} L. H. Gunderson, C. R. Allen, C. S. Holling {\it Foundation of Ecological Resilience}, Island Press; None edition (2009).
\bibitem{resilience2} C. Folke {\it Global Environmental Change} {\bf 16} (3): 253-267 (2006).
\bibitem{murray} J. D. Murray, {\it Mathematical Biology} 2nd ed. (Springer, New York, 2003).
\bibitem{strogatz} S. H. Strogatz, {\it Sync: The Emerging Science of Spontaneous Order} (Penguin, 2004).
\bibitem{menck} P. J. Menck, J. Heitzig, N. Marwan, J. Kurths, {\it Nat. Phys.} {\bf 9} 89-92 (2013).
\bibitem{trefethen} L. N. Trefethen, M. Embree, {\it Spectra and Pseudospectra: The Behavior of Nonnormal Matrices and Operators} (Princeton University Press, Princeton, 2005).
\bibitem{teo1} M. Asllani, R. Lambiotte, T. Carletti, {\it Sci. Adv.} {\bf 4} eaau9403 (2018).
\bibitem{zagli} D. Fanelli, F. Ginelli, R. Livi, N. Zagli, and C. Zankoc, {\it Phys. Rev E} {\bf 96} 062313 (2017).
\bibitem{sara} S. Nicoletti, N. Zagli, D. Fanelli, R. Livi, T. Carletti, G. Innocenti, {\it Phys. Rev. E} {\bf 98} 032214 (2018).
\bibitem{othmer1} H. G. Othmer and L. E. Scriven, {\it J. Theor. Biol.} {\bf 32}, 507 (1971).
\bibitem{othmer2} H. G. Othmer and L. E. Scriven, {\it J. Theor. Biol.} {\bf 43}, 83 (1974).
\bibitem{Mikhailov} H. Nakao and A. S. Mikhailov, {\it Nat. Phys.} {\bf}, 544 (2010).
\bibitem{Nakao} H. Nakao, {\it Eur. Phys. J. Special Topics} {\bf 223}, 2411 (2014).
\bibitem{ACPSF} M. Asllani, J. D. Challenger, F. S. Pavone, L. Sacconi, and D. Fanelli, {\it Nat. Commun.} {\bf 5}, 4517 (2014).
\bibitem{HNM} S. Hata, H. Nakao, and A. S. Mikhailov, {\it Sci. Rep.} {\bf 4}, 3585 (2014).
\bibitem{ABCFP} M. Asllani, D. M. Busiello, T. Carletti, D. Fanelli, and G. Planchon, {\it Phys. Rev. E} {\bf 90}, 042814 (2014).
\bibitem{Kouv} N. E. Kouvaris, S. Hata, and A. Diaz-Guilera, {\it Sci. Rep.} {\bf 5}, 10840 (2015).
\bibitem{PFMC} F. Di Patti, D. Fanelli, F. Miele, and T. Carletti, {\it Chaos Solitons Fractals} {\bf 96}, 8 (2017).
\bibitem{CBF} J. D. Challenger, R. Burioni, and D. Fanelli, {\it Phys. Rev. E} {\bf 92}, 022818 (2015).
\bibitem{PLFC} J. Petit, B. Lauwens, D. Fanelli, and T. Carletti, {Phys. Rev. Lett.} {\bf 119}, 148301 (2017).
\bibitem{LFC} M. Lucas, D. Fanelli, T. Carletti, and J. Petit, {\it Europhys. Lett.} {\bf 121}, 50008 (2018).
\bibitem{teo2} M. Asllani, T. Carletti, {\it Phys. Rev. E} {\bf 97} 042302 (2018).
\bibitem{muolo} R. Muolo, M. Asllani, D. Fanelli, P. K. Maini, T. Carletti {\it arXiv preprint arXiv: 1812.02514} (2019).
\bibitem{gardiner} C. W. Gardiner, {\it Handbook of Stochastic Methods} (Springer, Berlin, 2004).
\bibitem{vankampen} N. G. van Kampen, {\it Stochastic Processes in Physics and Chemistry} 3rd ed. (Elsevier, Amsterdam, 2007).
\bibitem{biancalani} T. Biancalani, F. Fafarpour, N. Goldenfeld, {\it Phys. Rev. Lett.} {\bf 118} 018101 (2017).
\bibitem{clement} C. Zankoc, D. Fanelli, F. Ginelli, R. Livi, {\it Phys. rev. E} {\bf 99} 012303 (2019).
\bibitem{chimera1} D. M. Abrams, S. H. Strogatz, {\it Phys. Rev. Lett.} {\bf 93} (2004).
\bibitem{chimera2} E. Sch\"oll, {\it Eur. Phys. J. Spec. Top.} {\bf 225} 891-919 (2016).
\bibitem{chimera3} M. J. Panaggio, D. M. Abrams, {\it Nonlinearity} {\bf 28} R67 (2015).
\bibitem{neubert} M. Neubert, H. Caswell,{\it Ecology} {\bf 78(3)} 653-665 (1997).
\bibitem{asllani} M. Asllani, J. D. Challenger, F. S. Pavone, L. Sacconi, D. Fanelli, {\it Nat. Commun.}  5:4517 (2014).
\bibitem{mckane} M. Asllani, T. Biancalani, D. Fanelli, A. J. McKane, {\it Eur. Phys. J. B} 86:476 (2013).
\bibitem{dipatti} F. Di Patti, D. Fanelli, F. Miele, T. Carletti, {\it CNSNS} {\bf 56} 447-456 (2018).
\bibitem{dipatti2} M. Asllani, F. Di Patti, D. Fanelli, {\it Phys. Rev. E} {\bf 86} 046105 (2012).
\bibitem{cencetti} G. Cencetti, F. Bagnoli, G. Battistelli, L. Chisci, F. Di Patti, D. Fanelli, {\it Eur. Phys. J. B} 90:9 (2017).
\bibitem{planchon} M. Asllani, D. M. Busiello, T. Carletti, D. Fanelli, G. Planchon, {\it Phys. Rev. E} {\bf 90} 042814 (2014).
\bibitem{deissler} R. J. Deissler, K. Kaneko, {\it Phys. Lett. A} {\bf 119} 397 (1987).
\bibitem{lepri1} S. Lepri, A. Politi, A. Torcini, {\it Phys. Lett.} {\bf 82} 1429 (1996).
\bibitem{lepri2}  S. Lepri, A. Politi, A. Torcini, {\it J. Stat. Phys.} {\bf 88} 31 (1997).
\bibitem{jiotsa} A. K. Jiotsa, A. Politi, A. Torcini, {\it J. Phys. A} {\bf 46} 254013 (2013).
\end{thebibliography}
\end{document}